\documentclass[twoside,10pt]{article}
\usepackage{arXiv_jmlr2e}
\usepackage{notation}
\usepackage{xcolor}
\usepackage{hyperref}
\hypersetup{colorlinks=true,linkcolor=violet,citecolor=violet,urlcolor=violet}

\usepackage{graphicx}
\usepackage{amsmath}
\usepackage{cleveref}
\usepackage{mathtools}
\usepackage{xspace}
\usepackage{multicol}
\usepackage{multirow}
\usepackage{booktabs}
\usepackage{caption}
\usepackage{algorithm}
\usepackage{algorithmic}

\def \xijt {{\mathbf{x}_{ij}^t}}

\def \xijs {{\mathbf{x}_{ij}^s}}

\def \yijs {{y_{ij}^s}}

\def \gbijsz {\gb_{ij}^{s, 0}}

\def \tp  {t^\prime}
\def \ijp {i^\prime, j^\prime}

\def \btheta {\boldsymbol{\theta}}
\def \bTheta {\boldsymbol{\Theta}}
\def \bR {\mathbb{R}}

\def \bSigma {\mathbf{\Sigma}}
\def \bigO {\mathcal{O}}

\newcommand{\la}{\langle}
\newcommand{\ra}{\rangle}

\newcommand{\model}{{P$^2$NeurRank}}
\newcommand{\linearmodel}{{P$^2$LinRank}}

\jmlrheading{1}{2000}{1-48}{4/00}{10/00}{Yiling Jia, Hongning Wang}

\ShortHeadings{Scalable Exploration for Neural Online Learning to Rank with Perturbed Feedback}{Jia and Wang}
\firstpageno{1}

\begin{document}
% \fancyhead{}
\title{Scalable Exploration for Neural Online Learning to Rank with Perturbed Feedback}

\author{\name Yiling Jia \email yj9xs@virginia.edu \\
	\addr Department of Computer Science\\
	University of Virginia\\
	Charlottesville, VA 22903, USA
	\AND
	\name Hongning Wang \email hw5x@virginia.edu\\
	\addr Department of Computer Science\\
	University of Virginia\\
	Charlottesville, VA 22903, USA
}
\maketitle

\begin{abstract}

Deep neural networks (DNNs) demonstrate significant advantages in improving ranking performance in retrieval tasks. 
Driven by the recent technical developments in optimization and generalization of DNNs, learning a neural ranking model online from its interactions with users becomes possible.
However, the required exploration for model learning has to be performed in the entire neural network parameter space, which is prohibitively expensive and limits the application of such online solutions in practice.

In this work, we propose an efficient exploration strategy for online interactive neural ranker learning based on the idea of bootstrapping. Our solution employs an ensemble of ranking models trained with perturbed user click feedback. The proposed method eliminates explicit confidence set construction and the associated computational overhead, which enables the online neural rankers' training to be efficiently executed in practice with theoretical guarantees. 
Extensive comparisons with an array of state-of-the-art OL2R algorithms on two public learning to rank benchmark datasets demonstrate the effectiveness and computational efficiency of our proposed neural OL2R solution.
\end{abstract}

\begin{keywords}
Online learning to rank; perturbed exploration; regret analysis
\end{keywords}

\section{Introduction}
\label{sec:intro}

Online learning to rank (OL2R) has recently attracted great research interest because of its unique advantages in capturing users' ranking preferences without requiring expensive relevance labeling as in classical offline learning to rank solutions~\citep{yue2009interactively,schuth2016multileave,wang2019variance,jia2021pairrank,lattimore2018toprank,li2018online,kveton2018bubblerank}. Because users' implicit feedback is noisy and biased~\citep{joachims2005accurately,agichtein2006improving,joachims2007evaluating,chapelle2012large}, the key in OL2R is to effectively explore the unknowns for improved relevance estimation, while serving the users with high-quality ranked results, which is known as the explore-exploit trade-off.

Most existing work in OL2R assumes a linear scoring function \citep{yue2009interactively,schuth2016multileave,wang2019variance}. Dueling bandit gradient descent (DBGD) and its different variants are the most popularly used OL2R solutions~\citep{yue2009interactively}, where new model variants are sampled via random perturbations in the parameter space to estimate the direction for model update.
As its non-linear extension, pairwise differential gradient descent (PDGD)~\citep{oosterhuis2018differentiable} samples the next ranked document from a Plackett-Luce model and estimates an unbiased gradient from the inferred pairwise preference. PairRank~\citep{jia2021pairrank} learns a logistic ranker online in a pairwise manner and explores the ranking space based on the model's estimation uncertainty about the pairwise comparisons of document rankings. Though practically effective, such a linear or generalized linear model assumption is incompetent to capture the possible complex non-linear relations between a document's ranking features and its relevance quality. This is already proved to be crucial in the past offline learning to rank practices~\citep{liu2011learning,burges2010ranknet}.

To unleash the power of representation learning, deep neural networks (DNN) have been introduced to learn the underlying scoring function for document ranking. In~\citep{oosterhuis2018differentiable}, PDGD is also experimented on a neural ranker. Though the authors reported promising empirical results, its theoretical property (e.g., convergence) is unknown.
On the other hand, enabled by the substantial progress in optimization and generalization of DNNs, quantifying a neural model's uncertainty on new data points become possible~\citep{cao2019generalization1, cao2019generalization2, chen2019much, daniely2017sgd, arora2019exact}. A recent work named olRankNet~\citep{jia2022neural} extended PairRank with a neural network ranker, which performs exploration in the pairwise document ranking space with topological sort by using the neural tangent kernel technique~\citep{jacot2018neural}. Compared with PairRank, olRankNet provided encouraging performance improvement, which was reported to be the best among all state-of-the-art OL2R solutions. More importantly, olRankNet is proved to achieve a sublinear gap-dependent regret upper bound, which is defined on the total number of mis-ordered pairs over the course of interactions with users. To our best knowledge, olRankNet is the first known OL2R solution for neural rankers with theoretical guarantees. 

% disucss the efficiency problem in online neural algorithms and disucss the perturbation based models, and also why olranknet cannot directly use the perturbation idea.
Despite being theoretically sound and empirically effective, olRankNet's limitation is also remarkably serious: its computational cost for performing the required exploration is prohibitively high (almost \emph{cubic} to the number of neural network's parameters). More specifically, to quantify the uncertainty of its estimated pairwise preferences among candidate documents, it has to maintain a high-probability confidence set for the current ranker's parameter estimation over time. However, the construction of the confidence set depends on the dimensionality of the neural network's parameters: as required by the neural tangent kernel, an inverse of the covariance matrix computed based on the gradient of the entire neural network is needed whenever the network is updated. 
For example, for a simple two layer feed-forward neural network with input dimension $d$ and $m$ neurons in each layer, the size of the covariance matrix is $(md + m^2+m)^2$. The best known time complexity for computing the inverse of this covariance matrix is $O\big((md + m^2+m)^{2.373}\big)$, by the optimized Coppersmith–Winograd algorithm \citep{williams2012multiplying}. 
This computational complexity quickly outpaces the limit of any modern computational machinery, given $m$ or $d$ are usually very large in practice (e.g., $m$ is often in the hundreds and $d$ in tens of thousands) and such matrix inverse operation is needed in every round when the neural network is updated. 
%often has the number of parameters in the order of 100 thousands. It is prohibitively expensive to compute the inverse covariance matrix on such a huge number of parameters, which is required for the confidence set construction. 
Due to this limitation, olRankNet has to employ the diagonal approximation of the covariance matrix in its actual implementations~\citep{jia2022neural}. But such an approximation loses its all theoretical guarantees, which unfortunately leads to a gap between the theoretical and empirical performance of olRankNet. And even how this gap would depend on the dimensionality of the network and affect olRankNet's performance is completely unknown. This inevitably limits the application of the neural OL2R solutions, especially the olRankNet-type algorithms, in practice.

In this work, we develop an efficient and scalable exploration strategy for olRankNet by eliminating its explicit confidence set construction. The basic idea is to use bootstrapping technique to measure the uncertainty of a neural ranker's output via a set of sample estimates. In particular, we maintain $N$ rankers in parallel. And in each round, after receiving the user's click feedback, each of the rankers is updated with the observed clicks and independently generated pseudo noise from a zero-mean Gaussian distribution. The overall model's estimation uncertainty on a pair of documents is then determined by an ensemble of the estimates from all $N$ rankers. For example, for a document pair $(i, j)$, if all $N$ rankers predict $i \succ j$, $(i, j)$ is considered as in a certain rank order, otherwise it is considered as in an uncertain rank order, where exploration is needed. Besides regular neural network updates, no additional computation is needed, which greatly reduces the computational overhead as required in olRankNet. We name our new solution as Perturbed Pairwise Neural Rank (or \model{} in short).
We rigorously prove that with a high probability \model{} obtains the same regret as olRankNet, but the computational complexity is way much lower. In addition, as no approximation is needed in \model{}, its theoretical analysis directly suggests its empirical performance.
Our extensive empirical evaluations demonstrate the strong advantage in both efficiency and effectiveness of \model{} against olRankNet and a rich set of state-of-the-art solutions over two OL2R benchmark datasets on standard retrieval metrics.  
\section{Related Work}

\noindent\textbf{Online learning to rank with neural rankers.} Most of the existing parametric OL2R solutions are limited to linear ranking models \citep{yue2009interactively,li2018online}. In particular, DBGD and its extensions \citep{yue2009interactively,schuth2014multileaved,wang2019variance,wang2018efficient}, as the most popularly referred OL2R solutions, are inherently designed for linear models as they rely on random perturbations in linear model weights for parameter estimation. But such a linear assumption is incompetent to capture any non-linear relations about documents' relevance quality under given queries, which shields such OL2R algorithms away from the successful practices in offline learning to rank models that are empowered by DNNs~\citep{burges2010ranknet,pasumarthi2019tf}. 

Such a limitation motivates some preliminary attempts in OL2R. In~\citep{oosterhuis2018differentiable}, pairwise differentiable gradient descent (PDGD) is proposed to sample the next ranked document from a Plackett-Luce model and estimate an unbiased gradient from the inferred pairwise ranking preference. Although improved empirical performance is reported for PDGD with a neural ranker, there is no theoretical guarantee on its online performance. Taking a completely different perspective, PairRank~\citep{jia2021pairrank} directly learns a pairwise logistic regression ranker online and explores the pairwise ranking space via a divide-and-conquer strategy based on the model's uncertainty about the documents' rankings. The authors claimed logistic regression can be treated as a one-layer feed-forward neural network, but it is unclear how PairRank can be extended to more general neural ranking architectures.

Recently, substantial progress in optimization and generalization of DNNs enables theoretical analysis about the neural models~\citep{liang2016deep, telgarsky2015representation, telgarsky2016benefits, yarotsky2017error, yarotsky2018optimal, lu2017depth, hanin2017approximating, zou2018stochastic, zou2019improved}. For example, with the neural tangent kernel technique~\citep{jacot2018neural}, the uncertainty of a neural model's estimation on new data points can be quantified~\citep{zhou2019neural, zhang2020neural, jia2022perturbation}. Most recently, olRankNet~\citep{jia2022neural} is proposed to extend PairRank with a multi-layer neural ranker. The authors proved that the good theoretical properties of PairRank (i.e., sublinear regret) are inherited in olRankNet, and empirically improved performance over PairRank was also reported in olRankNet. However, one serious issue of olRankNet is its cumbersome computational complexity: to quantify the confidence interval of the neural ranker's output for the exploration purpose, one has to compute the inverse of a covariance matrix which is derived by the gradient of entire neural network. The complexity is almost cubic to the number of parameters in the neural network, which is prohibitively expensive for OL2R, as this matrix inverse is needed every time the ranker is updated. The authors in~\citep{jia2022neural} suggested using diagonal approximation for the covariance matrix, but no guarantee is provided about the impact of such an approximation. 
%Besides, olRankNet is proved to achieve a sublinear regret upper bound, which is the first OL2R solution for a neural ranker with theoretical guarantee.

\noindent\textbf{Randomized exploration in online learning.} Efficient exploration is critical for online algorithms, as the model learns by actively acquiring feedback from the environment \citep{lattimore19bandit}. Distinct from the deterministic exploration strategies, such as upper confidence bound \citep{auer2002using,abbasi2011improved}, randomization-based exploration enjoys advantages in its light computational overhead and thus has received increasing attention in online learning community. The most straightforward randomization-based exploration strategy is $\epsilon$-greedy \citep{auer2002using}, which takes the currently estimated best action with probability $1-\epsilon$, otherwise randomly take an action. It has been applied in OL2R in \citep{hofmann2013balancing}. Almost no additional computation is needed in $\epsilon$-greedy for exploration, but the exploration is also independent from the current model estimation and therefore can hardly be optimal in practice. 
More advanced randomization-based exploration strategies are built on the bootstrapping technique in statistics. Giro~\citep{kveton2019garbage} explores by updating a model with a bootstrapped sample of its history with pseudo reward. In~\citep{kveton19perturbed,kveton20randomized}, random noise is added to the observed feedback for the model training to achieve the purpose of exploration in model's output. Such a strategy is proved to be effective in both linear and generalized linear model training. Most recently, Jia et al. \citep{jia2022perturbation} proved randomization can also be used for online neural network learning. The most closely related work to our study is~\citep{ash2021anti, ishfaq2021randomized}, where an ensemble of models are trained to approximate the confidence interval for the purpose of exploration in online model update.
\section{Method}
In this section, we provide a brief introduction of the general problem setting in OL2R, and then present our proposed solution for scalable exploration in neural OL2R.

\label{sec:method}
\subsection{Problem Formulation}

In OL2R, a ranker directly learns from the interactions with users for $T$ rounds. At each round $t=1,...,T$, the ranker receives a query $q_t$ and its associated $V_t$ candidate documents represented as a set of $d$-dimensional query-document feature vectors, $\mathcal{X}_t = \{\bx_1^t, \bx_2^t, \dots \bx_{V_t}^t\}$ with $\bx^t_i \in \bR^d$, and we assume that $\Vert\bx_i^t\Vert \leq u$. Once the query is received, the ranker determines the ranking of the $V_t$ documents based on its knowledge about the documents' relevance so far. We denote $\pi_t = (\pi_t(1), ..., \pi_t(V_t)) \in \Pi([V_t])$ as the ranking of the $L_t$ documents, where $\pi_t(i)$ represents the rank position of document $i$ given query $q_t$, and $\Pi([V_t])$ is the set of all permutations of the $V_t$ documents. After the ranked list is returned to the user, the user examines the list and provides her click feedback $C_t = \{c_1^t, c_2^t, ..., c_{V_t}^t\}$, where $c_i^t = 1$ if the user clicks on document $i$ at round $t$; otherwise $c_i^t = 0$. The ranker updates itself according to the feedback and proceeds to the next query.

Existing studies have repeatedly demonstrated that $C_t$ is biased and noisy~\citep{joachims2005accurately,agichtein2006improving,joachims2007evaluating,chapelle2012large}. Users tend to click more on the top-ranked documents, which is known as the \emph{position bias}; and as users can only interact with the documents shown to them, the ranker only has partial observations about relevance feedback from user clicks, which is known as the \emph{presentation bias}. Therefore, a good OL2R solution needs to carefully deal with the biased implicit feedback and effectively explore the unknowns for improved relevance estimation on the one hand, and serve users with the currently best estimated ranked result on the other hand. 

In this work, we follow the standard practice and treat clicks as relative preference feedback \citep{joachims2005accurately}. More specifically, the clicked documents are assumed to be preferred over those examined but unclicked documents. Besides, we consider every document that precedes a clicked document and the first $l$ subsequent unclicked document as examined. Such an assumption is widely adopted and proved to be effective in both offline and online learning to rank~\citep{wang2019variance, agichtein2006improving, oosterhuis2018differentiable, jia2021pairrank}. In particular, we denote $o_t$ as the index of the last examined position in the ranked list $\pi_t$ at round $t$.

Different from offline learning to rank, OL2R needs to serve the users while learning from its presented rankings. Therefore cumulative regret is an important metric for evaluating OL2R. In this work, we follow the regret defined as the number of mis-ordered pairs from the presented ranking to the ideal one~\citep{jia2021pairrank,lattimore2018toprank,jia2022neural},
\begin{equation*}
    R_T = \mathbb{E}\big[\sum\nolimits_{t=1}^T r_t\big] = \mathbb{E} \big[\sum\nolimits_{t=1}^T K(\pi_t, \pi_t^*)\big]
\end{equation*}
where $\pi_t$ is the ranked list generated by the current ranker, $\pi_t^*$ is the optimal ranking for the current query, and $K(\pi_t, \pi_t^*)=\Big|\big\{(i,j):i<j,\big(\pi_{t}(i)<\pi _{t}(j)\wedge \pi^*_{t}(i)>\pi^*_{t}(j)\big)\vee \big(\pi _{t}(i)>\tau _{t}(j)\wedge \pi^*_{t}(i)<\pi^*_{t}(j)\big)\big\}\Big|$. Such a pairwise regret definition directly connects an OL2R algorithm's online performance with classical ranking evaluations as most ranking metrics, such as ARP and NDCG can be decomposed into pairwise comparisons \citep{Wang2018Lambdaloss}.

\subsection{Exploration in olRankNet}

In the recent decade, DNNs have demonstrated powerful representation learning capacity and significantly boosted the performance for a wide variety of machine learning tasks \citep{goodfellow2016deep,lecun2015deep}, including learning to rank \citep{huang2013learning,dehghani2017neural,guo2020deep,severyn2015learning}. OL2R, on the other hand, has received limited benefit from the advances in DNNs. While DNNs are generally more accurate at predicting a document's relevance under a given query (i.e., exploitation), creating practical strategies to balance exploration and exploitation for neural ranking models in sophisticated online learning scenarios is challenging.

Built on the advances in renewed understandings about the generalization of DNNs, olRankNet extends PairRank with a neural scoring function and demonstrates the best empirical performance with theoretical guarantees~\citep{jia2022neural}. olRankNet directly learns a neural ranker from users' implicit feedback with a fully connected neural network $f(\xb;\btheta) = \sqrt{m}\Wb_L \phi(\Wb_{L-1} \phi(\dots \phi(\Wb_1\xb)))$, where depth $L \geq 2$, $\phi(\xb) = \max\{\xb, 0\}$, and $\Wb_1 \in \RR^{m \times d}$, $\Wb_i \in \RR^{m \times m}$, $2\leq i \leq L-1$, $\Wb_L \in \RR^{m \times 1}$, and $\btheta = [\text{vec}(\Wb_1)^\top,\dots,\text{vec}(\Wb_L)^\top]^\top \in \RR^{p}$ with $p = m+md+m^2 (L-2)$. At each round, the model $\btheta_t$ is updated by optimizing the cross-entropy loss between the predicted pairwise relevance distribution on all documents and those inferred from user feedback till round $t$ with a $\ell_2$-regularization term centered at the randomly initialized parameter $\btheta_0$:
\begin{align}
\label{eq:loss}
   \cL_t(\btheta) = \sum\nolimits_{s=1}^t\sum\nolimits_{(i, j) \in \Omega_s} -(1 - \yijs)\log\big(1 - \sigma(f_{ij})\big) -  \yijs\log\big(\sigma(f_{ij})\big) + {m \lambda}/{2}\|\btheta - \btheta_0\|^2,
\end{align}
where $f_{ij}^t$ = $f(\xb_i;\btheta_{t-1}) - f(\xb_j; \btheta_{t-1})$ is the difference between the estimated ranking scores of document $i$ and $j$, $\lambda$ is the $\ell_2$-regularization coefficient, $\Omega_s$ denotes the set of document pairs that received different click feedback at round $s$, i.e. $\Omega_s = \{(i, j): c_i^s \neq c_j^s, \forall \tau_s(i) \leq \tau_s(j) \leq o_t\}$, $\yijs$ indicates whether document $i$ is preferred over document $j$ based on the click feedback, i.e., $\yijs = (c_i^s - c_j^s+1)/2$~\citep{burges2010ranknet}. 

However, there is uncertainty in the estimated model $\btheta_t$ due to the click noise, i.e., $\Vert{\btheta}_t - \btheta^*\Vert \neq 0$, where $\btheta^*$ is assumed to be the underlying ground-truth model parameter. And therefore the model's output ranking might be wrong because of this uncertainty. olRankNet decides to randomize its output document rankings where its estimation is still uncertain, which helps collect unbiased feedback for improved model estimation subsequently.
Based on the neural tangent kernel technique, the uncertainty of olRankNet's estimated pairwise rank order can be analytically quantified and upper bounded with a high probability, under the assumption that pairwise click noise follows a $R$-sub-distribution \citep{jia2022neural}. This is described in the following lemma.

\begin{lemma}(Confidence Interval of Pairwise Rank Order in olRankNet). 
\label{lemma_CI}
There exist positive constants $C_1$ and $C_2$ such that for any $\delta_1 \in (0,1)$, with satisfied step size of gradient descent $\eta$, and the neural network width $m$, at round $t < T$, for any document pair $(i, j)$ under query $q_t$, with probability at least $1 - \delta_1$,
\begin{equation}
\label{eq:cb}
    |\sigma(f_{ij}^t) - \sigma(f_{ij}^*) | \leq \alpha_t\Vert\gb^t_{ij}/\sqrt{m}\Vert_{\mathbf{\Ab}_t^{-1}} + \epsilon(m),
\end{equation}
where $\epsilon(m)$ is the approximation error from gradient descent in neural network optimization,  $f_{ij}^*$ is the ground-truth difference between document pair $(i, j)$, $\gb_{ij}^s = \gb(\xb_i; \btheta_s) - \gb(\xb_j; \btheta_s)$ with $\gb(\xb; \btheta)$ as the gradient of input $\xb$ with respect to the entire network parameters, $\alpha_t = \bar C_1\Big(\sqrt{R^2\log ({\det(\Ab_t)}/{\delta_1^2 \det(\lambda \Ib))}} + \sqrt{\lambda}{\bar C_2}\Big)$ with $\bar C_1$ and $\bar C_2$ as positive constants, $\Ab_t = \sum_{s=1}^{t-1}\sum_{(i^\prime, j^\prime) \in \Omega_{s}}
\frac{1}{m}\gb_{\ijp}^s{\gb_{\ijp}^s}^\top + \lambda \mathbf{I}$.
\label{lemma:pairrank_cb}
\end{lemma}

With the constructed confidence interval of the estimated pairwise document rank order, olRankNet separates all the candidate document pairs into two sets, certain rank order $S_t^c$ and uncertain rank order $S_t^u$. $S_t^c$ contains the document pairs where with high probability the estimated pairwise order is correct. For example, for document $i$ and $j$, if the lower confidence bound of the probability that $i$ is better than $j$, i.e., $\sigma(f_{ij}^t) - \alpha_t\|\gb_{ij}^t/\sqrt{m}\|_{\Ab_t^{-1}} - \epsilon(m)$, is larger than 0.5, then $(i, j)$ belongs to $S_t^c$. Otherwise, the pair belongs to $S_t^u$, which indicates that the predicted rank order can still be wrong.

When constructing the ranked list, olRankNet first builds a ranking graph with all the candidate documents and the certain rank orders in $S_t^c$. Then topological sort is performed, where the certain rank orders will be followed (i.e., exploitation), and uncertain rank orders will be randomized (i.e., exploration). With such a ranking strategy, olRankNet is proved to have an $\bigO{(\log^2(T))}$ cumulative pairwise regret upper bound.

Although olRankNet has a strong theoretical foundation, its scalability is severely limited due to the additional computation required for constructing the confidence interval. In particular, the covariance matrix $\Ab$ in Lemma~\ref{lemma:pairrank_cb} is constructed with the gradient of the scoring function with respect to the network parameters, of which the size $p$ is very large. In order to construct the confidence interval according to Eq~\eqref{eq:cb}, the inverse of the covariance matrix $\Ab$ has to be computed whenever the model is updated, which results in an unacceptably high computational cost (around $O(p^3)$). As a consequence, it is practically impossible for olRankNet to be exactly executed. In~\citep{jia2022neural}, approximation is employed to make olRankNet operational in practice, e.g., only using the diagonal of $\Ab$. However, there is no theoretical guarantee for such an approximation, which unfortunately breaks the theoretical promise of olRankNet and directly leads to an unknown gap between its theoretical and empirical performance.

\subsection{Scalable Exploration with Perturbed Feedback}

\begin{figure}[t]
    \centering
    \includegraphics[width=0.7\linewidth]{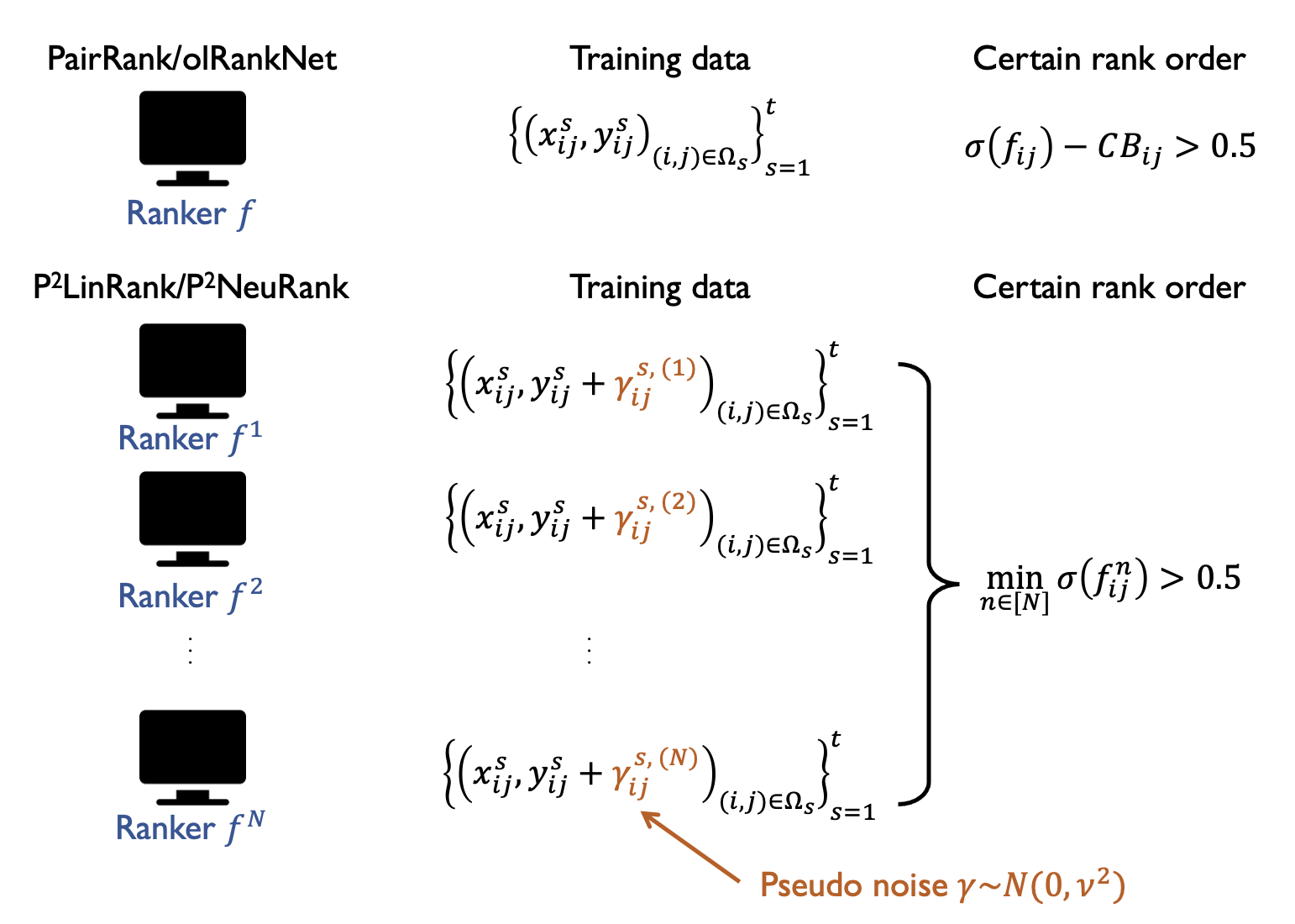}
    \caption{Comparison between PairRank/olRankNet and P$^2$LinRank/P$^2$NeuRank for the procedures in model training and result ranking.}
    \label{fig:compare}
\end{figure}

To bridge the gap, we develop an efficient and scalable strategy for recognizing the certain and uncertain rank orders without explicitly constructing the confidence set. And our basic idea is to leverage the bootstrapping technique to create randomness in a neural ranker's output. In particular, at each round, we perturb the entire user feedback history for $N$ times with noise freshly and independently sampled from a zero-mean Gaussian distribution, and train the corresponding neural ranker as usual. Denote model $\btheta^{(n)}$ for $n \in [N]$ as the solution of minimizing the following objective function with gradient descent, 
\begin{align}
\label{eq:obj}
   \btheta^{(n)} =& \min\sum\nolimits_{s=1}^t\sum\nolimits_{(i, j) \in \Omega_s} -\big(1 - (\yijs+\gamma_{ij}^{s, (n)})\big)\log\big(1 - \sigma(f_{ij})\big) \nonumber \\
     &-  (\yijs + \gamma_{ij}^{s, (n)})\log\big(\sigma(f_{ij})\big) + {m \lambda}/{2}\|\btheta - \btheta_0\|^2,
\end{align}
where $\{\gamma_{ij}^{s, (n)}\}_{s=1}^t \sim \cN(0, \nu^2)$ are Gaussian random variables that are independently sampled in each round $t$, and $\nu$ is a hyper-parameter that controls the strength of perturbation (and thus the exploration) in \model{}.

The detailed procedure of \model{} is given in Algorithm~\ref{alg:model}. The algorithm starts by initializing the $N$ neural rankers. At each round of interaction, given a query $q_t$, for each pair of candidate documents, $N$ parallel predictions about their rank order will be generated by the set of neural rankers. If all the $N$ estimations give the same prediction about the document pair's rank order, e.g., $i \succ j$ for document $i$ and document $j$, then $(i, j)$ is considered as a certain rank order (line 9 - line 13 in Algorithm~\ref{alg:model}). Otherwise, the relation between these two documents is still uncertain and further exploration is needed there when generating the ranked list. Once the sets of certain and uncertain rank orders are determined, we follow the same procedure of olRankNet to generate the ranked list via topological sort with respect to the certain rank orders.

The key intuition for \model{} is to utilize the variance introduced in the randomly perturbed click feedback to encourage exploration. With the injected perturbation, there are two kinds of deviations existing in the estimated pairwise preference in each of the $N$ parallel neural rankers: 1) the deviation caused by the observation noise introduced by the click feedback; 2) the deviation caused by the added perturbations. By properly setting the variance parameter $\nu$ for the added perturbation, the corresponding deviation will introduce enough randomness in the model estimation. For each round of interaction, we maintain $N$ models and with high probability, the minimum of the estimated pairwise preference serves as the pessimistic estimate of the preference.

Compared to olRankNet, which requires to maintain the inverse of the covariance matrix, \model{} does not need any added computation for the purpose of exploration, besides the regular neural network updates. As a result, \model{} greatly alleviates the computation burden in neural OL2R. More specifically, in each round, the online neural ranking algorithm generally takes the following steps: (1) predict the rank order in each document pair, (2) generate the ranked list by topology sort via the constructed certain and uncertain rank orders, and (3) update the model according to the newly received click feedback. With $p$ representing the total number of parameters in a neural ranker, olRankNet has the time complexity $\bigO{(V_tp + V_t^2)}$ for the first step. As the $N$ neural models in \model{} are independent from each other, the time complexity of \model{} in the first step is also $\bigO{(V_tp + V_t^2)}$ by executing the $N$ ranker's predictions in parallel. For the third step, again by the training the $N$ neural rankers in parallel using gradient descent, both \model{} and olRankNet have the time complexity of $\bigO{(\tau p \sum_{s=1}^t|\Omega_s|)}$ where $\tau$ is the number of epochs for training the neural network. 
The key difference lies in the second step. olRankNet requires the inverse of covariance matrix, which has the time complexity at least $\bigO{(p^{2.2373})}$. Besides, constructing the confidence interval for all the document pairs has the time complexity of $\bigO{(V_t^2p^2)}$. While for \model{}, finding the minimum of the $N$ predictions for all the document pairs costs $\bigO{(NV_t^2)}$. Once the certain and uncertain rank orders are determined, both algorithms require $\bigO{(V_t + E_t)}$ for the topological sort, where $E_t$ represents the number of certain rank orders and $E_t \leq V_t$. Therefore, for the second step, olRankNet has the total time complexity as $\bigO{(p^{2.2373} + V_t^2p^2 + V_t + E_t)} = \bigO{(p^{2.2373} + V_t^2p^2)}$, while \model{} has the time complexity as $\bigO{(NV_t^2 + V_t + E_t)} = \bigO{(NV_t^2)}$. As $p$ is oftentimes in the order of tens of thousands (if not less), \model{} greatly reduces the time required for performing exploration in neural OL2R. And also empirically, the number of parallel rankers $N$ in \model{} does not need to be large. For example, in our experiments, we found $N=2$ already led to promising performance of \model{} comparing to olRankNet.

We want to highlight that our proposed perturbation-based exploration strategy can also be applied to linear ranking models, e.g., PairRank \citep{jia2021pairrank}. The procedure is almost the same as described in Algorithm \ref{alg:model}, and so is its computational advantage in linear models, especially when the dimension of the feature vectors is large. In our experiments, we empirically evaluated our perturbation-based method in PairRank (named \linearmodel{}) and observed its expected performance and computational advantages.

\begin{algorithm}[t]
	\caption{\model{}} 
	\label{alg:model} 
	\begin{algorithmic}[1]
    \STATE  \textbf{Input:}  Number of rounds $T$, regularization coefficient $\lambda$, perturbation parameter $\nu$, network width $m$, network depth $L$, and number of rankers $N$.
    \STATE Initialize $N$ neural network models $\{\btheta_0^{n}\}_{n=1}^N$ with $m$ and $L$
    \FOR{t = 1, ..., T}
    \STATE $S_t^c = \emptyset, S_t^u = \emptyset$ 
    \STATE $q_t \leftarrow$ receive\_query(t)
    \STATE $\mathcal{X}_t = \{\bx_1^t, \bx_2^t, \dots \bx_{V_t}^t\} \leftarrow$ retrieve\_candidate\_documents($q_t$)
    \FOR{each document pair $(i, j) \in [V_t]^2$}
    \STATE $\{\sigma(f_{ij, t}^{n})\}_{n=1}^N \leftarrow$ get\_N\_estimations($\bx_i^t$, $\bx_j^t$, $\{\btheta_t^{n}\}_{n=1}^N$)
    \IF{$\min_{n\in[N]}\sigma(f_{ij, t}^{n}) > 0.5$ or $\max_{n\in[N]}\sigma(f_{ij, t}^{n}) < 0.5$} 
    \STATE $S_t^c \leftarrow S_t^c \cup (i, j)$
    \ELSE 
    \STATE $S_t^u \leftarrow S_t^u \cup (i, j)$
    \ENDIF
    \ENDFOR
    \STATE $\pi_t \leftarrow $ topological\_sort($S_t^c$, $S_t^u$)
    \STATE $C_t \leftarrow $ collect\_click\_feedback($\pi_t$)
    \STATE $\Omega_t, \{y_{ij}\}_{(i,j)\in\Omega_t} \leftarrow $ construct\_training\_data($C_t$)
    \FOR{$n=1, ..., N$}
    \STATE Generate $\{\{\gamma_{ij}\}_{(i,j)\in\Omega_s}\}_{s=1}^t \sim \cN(0, \nu^2)$
    \STATE Set $\btheta_t^n$ by the output of gradient descent for solving Eq~\eqref{eq:obj} with $\{\Omega_s\}_{s=1}^t$.
    \ENDFOR
    \ENDFOR
	\end{algorithmic}
\end{algorithm}

% \begin{definition}
% \label{def:certain}
% At any round $t < T$, the ranking order between document $i$ and $j$, denoted as $(i, j)$, is considered in a certain rank order if and only if $\min_{m\in[M]} \sigma(\xijt^\top\hat{\btheta}_t^{(m)}) > \frac{1}{2}$ or $\max_{m\in[M]} \sigma(\xijt^\top\hat{\btheta}_t^{(m)}) < \frac{1}{2}$.
% \end{definition}

\section{Regret Analysis}
\label{sec:regret}

In this section, we provide the regret analysis of the proposed exploration strategy. For better readibility, we present the analysis of a linear ranker. According to the anlaysis in~\citep{jia2022neural}, under the neural tangent technique and the convergence analysis of the gradient descent in neural network optimization, the linear analysis can be readily applied to the neural ranker. And we discuss the difference between the analysis between the linear ranker and neural ranker in the appendix. 

Follow the standard assumption in~\citep{jia2021pairrank, jia2022neural}, we assume that on the \emph{examined} documents where $\pi_t(i) \leq o_t$, the obtained feedback $C_t$ is independent from each other given the \emph{true relevance} of documents, so is their noise \citep{joachims2005accurately,guo2009click,guo2009efficient}. Therefore, the noise in the inferred preference pair becomes the sum of noise from the clicks in the two associated documents. And we also only use the independent pairs to construct $\Omega_t$ as suggested in PairRank and olRankNet. Thus, the pairwise noise satisfies the following proposition.

\begin{proposition}
\label{prop:pairwise}
For any $t \geq 1$, $\forall (i, j) \in \Omega_t$, the pairwise feedback follows $y_{ij}^t = \sigma\big(f(\xb_i; \btheta^*) - f(\xb_j; \btheta^*)\big) + \epsilon_{ij}^t$, where $\epsilon_{ij}^t$ satisfies that for all $\beta \in \RR$, $\EE\Big[\exp(\beta\epsilon_{ij}^t) | \big\{\{\epsilon^s_{\ijp}\}_{\ijp \in \Omega_s}\big\}_{s=1}^{t-1}, \Omega_{1:t-1}\Big] \leq \exp(\beta^2R^2)$, is an $R$-sub-Gaussian random variable.
\end{proposition}

To train a linear ranker, we have the scoring function $f(\xb; \btheta) = \xb^\top\btheta$. And we assume that $\|\xb\| \leq P$ and $\|\btheta\| \leq Q$. The loss function can be rewritten as,
\begin{align}
\label{eq:linloss}
%   \cL^{(n)}_t(\btheta)^ =& \frac{ \lambda}{2}\|\btheta\|^2 + \sum_{s=1}^t\sum_{(i, j) \in \Omega_s} -(\yijs + \gamma_{ij}^{s, (n)})\log(\sigma(\xijs^\top\btheta)) \nonumber \\
%      &- (1 - (\yijs + \gamma_{ij}^{s, (n)}))\log(1 - \sigma(\xijs^\top\btheta)),
     \cL^{(n)}_t(\btheta)^ = \sum_{s=-d+1}^t\sum_{(i, j) \in \Omega_s} -\big(\yijs + \gamma_{ij}^{s, (n)}\big)\log(\sigma(\xijs^\top\btheta)) 
     - \big(1 - (\yijs + \gamma_{ij}^{s, (n)})\big)\log\big(1 - \sigma(\xijs^\top\btheta)\big),
\end{align}
where $\xijs = \bx_i^s - \bx_j^s$ is the difference between the feature vectors of document $i$ and $j$, and $d$ is the dimension of the feature vectors. With $|\Omega_s| = 1$, $\xijs = \sqrt{\lambda}\mathbf{e_i}$, $\yijs = 0$ for $s \in [-d+1, 0]$, this loss function can be interpreted as adding $l_2$ regularization to the cross-entropy loss.

Given this objective function is log-convex with respect to $\btheta$, its solution $\hat{\btheta}_t^{(n)}$ of ranker $n$ for $n \in [N]$ is unique under the following estimation method at each round $t$,
\begin{align}
\label{eqn:gradient}
    \sum\nolimits_{s=-d+1}^{t-1}\sum\nolimits_{(i, j)\in\Omega_s} \left(\sigma({\xijs}^\top\btheta) - (\yijs + \gamma_{ij}^{s, (n)}) \right)\xijs+ \lambda\btheta = 0
\end{align}

Let $g_t(\btheta) = \sum_{s=-d+1}^{ t-1}\sum_{(i, j) \in \Omega_s}\sigma({\xijs}^\top\btheta)\xijs + \lambda\btheta$ be the invertible function such that the estimated parameter $\hat{\btheta}_t^{(n)}$ satisfies $g_t(\hat{\btheta}_t^{(n)}) = \sum_{s=-d+1}^{t-1}\sum_{(i, j)\in\Omega_s}(\yijs + \gamma_{ij}^{s, (n)})\xijs$. 
As discussed before, there are two kinds of deviations inside this estimation $\hat{\btheta}_t^{(n)}$. To analyze their effect in the estimation, we introduce an auxiliary solution $\bar\btheta_t$ for solving the linear objective function, which satisfies $g_t(\bar \btheta_t)  =  \sum\nolimits_{s=1}^{t-1}\sum\nolimits_{(i, j)\in\Omega_s}\yijs\xijs$. Then, for the two solutions $\hat \btheta_t^{(n)}$ and $\bar \btheta_t$, we have the following lemmas quantifying the deviations in the estimation. 

\begin{lemma} (Deviation from observation noise). At round $t < T$, for any pair of document $(\bx_i^t, \bx_j^t)$ under query $q_t$, with probability at least $1 - \delta$, we have,
\begin{align*}
    |\sigma(\xijt^\top\bar{\btheta}_t) - \sigma(\xijt^\top\btheta^*_t)| \leq \alpha_t \|\xijt\|_{\bA_t^{-1}},
\end{align*}
where $\alpha_t = (2k_{\mu}/c_{\mu})\big(\sqrt{R^2\log(\det(\bA_t)/(\delta^2\det(\lambda\mathbf{I})))} + d\big)$, $\bA_t = \lambda \mathbf{I} + \sum_{s=1}^{t-1}\sum_{(\ijp) \in \Omega_s}\bx_{\ijp}\bx_{\ijp}^\top$, $k_{\mu}$ is the Lipschitz constant of the sigmoid link function $\sigma$, $c_{\mu} = \inf_{\btheta \in \bTheta} \dot{\sigma}(\bx^\top\btheta)$, with $\dot{\sigma}$ as the first derivative of $\sigma$.
\label{lemma:cb}
\end{lemma}

Accordingly, we define $E_t$ as the success event at round $t$:
\begin{equation*}
    E_t = \big\{ \forall (i, j) \in [V_t]^2, |\sigma({\mathbf{x}_{ij}^t}^\top\bar{\btheta}_t) - \sigma({\mathbf{x}_{ij}^t}^\top\btheta^*) | \leq \alpha_t\Vert\mathbf{x}_{ij}^t\Vert_{\bA_t^{-1}}\big\}.
\end{equation*}
Intuitively, $E_t$ is the event that the auxiliary solution $\bar{\btheta}_t$ is ``close'' to the optimal model $\btheta^*$ at round $t$. 
% Next, we analyze the deviation from the added pseudo noise with the following lemma.

% \begin{lemma}
% At round $t < T$, for the solutions $\bar \btheta_t$ and $\hat \btheta_t$, with the added pseudo noise satisfing $\gamma \sim \mathcal{N}(0, \nu^2)$, for any document $\bx_i^t$ and $\bx_j^t$ under query $q_t$, we have $|\sigma(\xijt^\top\bar{\btheta_t}) - \sigma(\xijt^\top\hat{\btheta_t})| \leq \frac{k_{\mu}}{c_{\mu}}|U(\xijt)|$,
% where $U(\xijt) \sim \mathcal{N}(0, \nu^2\|\xijt\|^2_{\bA_t^{-1}})$
% \end{lemma}

% We also have the following lemma for event $E_2$,

% With these two lemmas, we can analyze the regret. 
As discussed before, we define the regret as the number of mis-ordered pairs. Therefore, the key step in regret analysis is to quantify the probability that an estimated preference is uncertain. According to Algorithm~\ref{alg:model}, the certain rank order in the perturbed pairwise ranker is defined as follows,
\begin{definition} (Certain Rank Order)
\label{def:certain}
At round $t$, the rank order between documents $(i, j) \in [V_t]^2$ belongs to the set of certain rank orders $\omega_t^c$ if and only if $\min_{n\in[N]}\sigma\left(f_{ij}^{t, (n)}\right) > \frac{1}{2}$ or $\max_{n\in[N]}\sigma\left(f_{ij}^{t, (n)}\right) < \frac{1}{2}$; otherwise, $(i, j) \in \omega_t^u$.
\end{definition}

According to the definition, and the deviations caused by the observation noise and the pseudo noise, we have the following lemma quantifying the probability of an estimation being uncertain.
\begin{lemma}
\label{lemma:uncertain}
There exist positive constants $c$, $C_1$ and $C_2$, that with 
$t^\prime = \frac{2k_{\mu}P}{c_{\mu}\Delta_{\min}}\Big(\sqrt{R^2\log(1/\delta)} + \sqrt{\lambda}Q\Big) +  \Big( \frac{C_1\sqrt{d} + C_2\sqrt{\log(1/\delta)} + (P^2Rk_{\mu})/(\sqrt{\lambda}c_{\mu}\Delta_{\min})}{\lambda_{\min}(\Sigma)}\Big)^2$, $\delta \in (0, 1)$, for round $t \geq t^\prime$, with probability at least $1 - \delta$, event $E_t$ holds with $\alpha_t$ defined in Lemma~\ref{lemma:cb}, with $N \geq \log{\delta}/\log(1 - \exp(-\beta^2)/(4\sqrt{\pi}\beta))$, where $\beta = \frac{k_\mu^2\alpha_t^2}{c_{\mu}^2\nu^2}$, for a document pair $(i, j)$ that $i \succ j$ for the given query, the probability that the estimated pairwise preference is uncertain is upper bounded as $\mathbb{P}\big((i, j) \in \omega_t^u\big) \leq \frac{2N\nu^2k_{\mu}^2\|\xijt\|^2_{\Ab_t^{-1}}}{c_{\mu}^2c^2\Delta_{\min}^2}$,
where $\Delta_{\min} = \min\limits_{t\in T, (i, j) \in [V_t]^2}| \sigma({\xijt}^\top\btheta^*) - \frac{1}{2}|$ representing the smallest gap of pairwise difference between any pair of documents associated to the same query over time (across all queries).
% $CB = \alpha_t\Vert\mathbf{x}_{ij}^t\Vert_{\bA_t^{-1}}$ with $\alpha_t$ defined in Lemma~\ref{lemma:cb} with the $\delta$.
\end{lemma}

The detailed proof is provided in the appendix. This lemma provides the upper bound of the probability that an estimated pairwise preference is uncertain. The key idea is to analyze the concentration and anti-concentration property of the deviation caused by the pseudo noise. In particular, the deviation caused by the pseudo noise $\gamma$, and the ensemble of $N$ rankers should be sufficiently large so that for document pairs $(i, j)$, the maximum estimated pairwise preference, $\max_{n\in[N]} \sigma(\xijt^\top\hat\btheta_t^{(n)})$ is optimism to trigger exploration. On the other hand, with more observations, the probability of being uncertain will be shrinking with the concentration property of the pseudo noise.

Following the assumption in~\citep{kveton2015combinatorial, jia2021pairrank, jia2022neural}, denote $p_{t}$ as the probability that the user examines all documents in $\tau_t$ at round $t$, and let $p^* = \min_{1\leq t \leq T} p_{t}$ be the minimal probability that all documents in a query are examined over time. The regret of the proposed model can be upper bounded as follows.

\begin{theorem}
\label{theorem}
Assume pairwise query-document feature vector $\xijt$ under query $q_t$, where $(i, j) \in [V_t]^2$ and $t \in [T]$, satisfies Proposition 1. With  $\delta \in (0, 1)$, with probability at least $1 - \delta$, the $T$-step regret of the proposed model is upper bounded by:
\begin{align*}
    R_T 
     \leq& R^\prime + \frac{1}{p^*}2dV_{\max}C\log\left(1 + \frac{o_{\max}TP^2}{2d\lambda}\right)
\end{align*}
where $R^{\prime} = \tp V_{\max}$, $V_{\max}$ represents the maximum number of document associated with the same query over time, and $\tp$ is defined in Lemma \ref{lemma:uncertain}, and $w = \sum_{s=\tp}^T \big({(V_{\max}^2 - 2V_{\max})P^2 }/{\lambda_{\min}(\bA_s)}\big)$, and 
% $t'$ satisfies Lemma 2, and 
By choosing $\delta_1 = \delta_2 = 1/T$, we have the expected regret at most $R_T \leq O(d\log^2(T))$.
\end{theorem}

We provide the detailed proof in the appendix. According to the pairwise exploration strategy, the regret only comes from the document pairs that are uncertain, e.g., random shuffling will be conducted to perform the exploration. With the quantified uncertain probability in Lemma~\ref{lemma:uncertain}, the pairwise regret can be upper bounded accordingly.

In neural rankers, the neural network approximation error should be considered in addition to the deviations caused by the noise. According to the analysis  in~\citep{jia2022perturbation}, the variance of the added noise should be set according to the deviations caused by both the observation noise and the approximation error. Based on the theoretical analysis in~\citep{jia2022neural}, by properly setting the width of the neural network and the step size of gradient descent, the model with a neural ranker will still have a sublinear regret.

% \begin{hproof}
% The detailed proof is provided in the appendix. Here, we only provide the key ideas behind our regret analysis.
% The regret is first decomposed into two parts: $R^\prime$ represents the regret when either $E_t$ or Lemma~\ref{lemma:uncertain} does not hold, in which the regret is out of our control, and we use the maximum number of pairs associated to a query over time, $L_{\text{max}}$ to compute the regret. The second part corresponds to the cases when both events happen. Then, the instantaneous regret at round $t$ can be bounded by
% \begin{align}
%     r_t = \mathbb{E} \big[K(\tau_t, \tau_t^*)\big] = \sum\nolimits_{i=1}^{d_t}\mathbb{E}\big[\frac{(N_i^t + 1)N_i^t}{2}\big] \leq \mathbb{E}\big[\frac{N_t(N_t + 1)}{2}\big]
% \end{align}
% where $N_i^t$ denotes the number of uncertain rank orders in block $\mathcal{B}_i^t$ at round $t$, and $N_t$ denotes the total number of uncertain rank orders.
% From the last inequality, it follows that in the worst case where the $N_t$ uncertain rank orders are placed into the same block and thus generate at most $({N_t^2 + N_t})/{2}$ mis-ordered pairs with random shuffling. This is because based on the blocks created by \model{}, with $N_t$ uncertain rank orders in one block, this block can at most have $N_t + 1$ documents. Then, the cumulative number of mis-ordered pairs can be bounded by the probability of observing uncertain rank orders in each round, which shrinks rapidly with more observations over time.
% \end{hproof}
\section{Experiment}
\label{sec:exp}
In this section, we empirically compare our proposed model with an extensive list of state-of-the-art OL2R algorithms on two large public learning to rank benchmark datasets.

\subsection{Experiment Setup}
\subsubsection{Dataset. } We experimented on the Yahoo! Learning to Rank Challenge dataset \citep{chapelle2011yahoo}, which consists of 292,921 queries and 709,877 documents represented by 700 ranking features, and MSLR-WEB10K \citep{qin2013introducing}, which contains 10,000 queries, each having 125 documents on average represented by 136 ranking features. Both datasets are labeled on a five-grade relevance scale: from not relevant (0) to perfectly relevant (4). These two datasets are the most popularly used in literature for evaluating OL2R algorithms. We followed the train/test/validation split provided in the datasets to perform the cross-validation to make our results comparable to previously reported results. 

\subsubsection{User interaction simulation.} A standard simulation setting for the user clicks is adopted in our experiment \citep{oosterhuis2018differentiable, jia2021pairrank}, which is the most popularly used procedure for OL2R evaluations. At each round, a query is uniformly sampled from the training set. The model will then generate a ranked list and return it to the user. Dependent click model (DCM)~\citep{guo2009efficient} is applied to simulate user behaviors, which assumes that the user will sequentially scan the ranked list and make click decisions on the examined documents. In DCM, the probabilities of clicking on a given document and stopping the subsequent examination are both conditioned on the document's true relevance label. 
Three different model configurations are employed in our experiments to represent three different types of users. The details are shown in Table \ref{table:click}. In particular, \textit{perfect} users will click on all relevant documents and do not stop browsing until the last returned document; \textit{navigational} users are very likely to click on the first encountered highly relevant document and stop there; and \textit{informational} users tend to examine more documents, but sometimes click on irrelevant documents, which contribute a significant amount of noise in the click feedback. To reflect the presentation bias, all the models only return the top 10 ranked results. 

\begin{table}[t]
  \caption{Configuration of simulated click models.}
  \label{table:click}
  \centering
  \begin{tabular}{cccccc|ccccc}
    \hline
                & \multicolumn{5}{c}{Click Probability} & \multicolumn{5}{c}{Stop Probability} \\
R & 0           & 1          & 2   &3 &4        & 0          & 1          & 2   &3 &4       \\ \hline
\textit{per}         & 0.0         & 0.2        & 0.4 &0.8 &1.0        & 0.0        & 0.0        & 0.0   &0.0 &0.0     \\
\textit{nav}    & 0.05   &0.3      & 0.5    &0.7    & 0.95       & 0.2       &0.3 & 0.5    &0.7    & 0.9        \\
\textit{inf}   & 0.4      &0.6   & 0.7   &0.8     & 0.9        & 0.1      &0.2  & 0.3     &0.4   & 0.5        \\ \hline
\end{tabular}
\end{table}

\subsubsection{Baselines.}  We list the OL2R solutions used for our empirical comparisons below. We performed the experiments on both linear and neural rankers to show the general effectiveness of our proposed exploration strategy. For convenience of reference, in the experiment result discussions, we name our solution applied to a linear ranker as \linearmodel{} 

\begin{itemize}
    \item \textbf{$\epsilon$-Greedy (linear and neural)} \citep{hofmann2013balancing}: At each rank position from top to bottom, it randomly samples an unranked document with probability $\epsilon$ or selects the next best document based on the currently learned ranker.
    \item \textbf{DBGD (linear and neural)} \citep{yue2009interactively}: DBGD uniformly samples a direction in the entire model space for exploration and model update. Its convergence is only proved for linear rankers, though empirically previous studies also applied it to neural rankers. 
    % \item \textbf{MGD (linear and neural)} \citep{schuth2014multileaved}: MGD uniformly samples multiple direction from the entire model space to reduce the variance in DBGD. Again, MGD's theoretical guarantee is only for linear rankers. 
    \item \textbf{PDGD (linear and neural)} \citep{oosterhuis2018differentiable}: PDGD samples the next ranked document from a Plackett-Luce model and estimates gradients from the inferred pairwise preferences based on the observed user feedback. The only known theoretical property about PDGD is its estimated gradient is unbiased, but how this unbiased gradient leads model training, i.e., its convergence, is still unknown. 
    \item \textbf{PairRank}~\citep{jia2021pairrank}: PairRank learns a pairwise logistic regression ranker online and explores by divide-and-conquer in the pairwise document ranking space. The training of PairRank is known to converge with a sublinear regret defined on the cumulative number of misordered document pairs.
    \item \textbf{PairRank-Diag}: This is a variant of PairRank where the diagonal approximation of its covariance matrix is used to calculate the required confidence interval. 
    \item \textbf{olRankNet}~\citep{jia2022neural}: olRankNet is an extension of PairRank with a neural ranker, where the confidence interval is constructed via the neural tangent kernel technique. It inherits good theoretical properties from PairRank.
    \item \textbf{olRankNet-Diag}: This is a variant of olRankNet, where the diagonal approximation of the covariance matrix is applied to calculate the confidence interval.
\end{itemize}

\subsubsection{Hyper-Parameter Tuning. } MSLR-WEB10K dataset is equally partitioned into five folds, while Yahoo Learning to Rank dataset is equally partitioned into two folds. We performed cross validation on each dataset. For each fold, the models are trained on the training set, and the hyper-parameters are tuned based on the performance on the validation set.

In our experiments, for all the neural rankers, a two layer neural network with width $m=100$ is applied. We did a grid search in olRankNet, PairRank, PairRank-Diag for its regularization parameter $\lambda$ over $\{10^{-i}\}_{i=1}^4$, exploration parameter $\alpha$ over $\{10^{-i}\}_{i=1}^4$. For parameter estimation in all neural rankers, we did a grid search for learning rate of gradient descent over $\{10^{-i}\}_{i=1}^3$. PairRank and PairRank-Diag are directly optimized with L-BFGS. The model update in PDGD and DBGD is based on the optimal settings in their original paper (i.e., the exploration step size and learning rate). 
The hyper-parameters for PDGD and DBGD are the  learning rate and the learning rate decay, for which we performed a grid search for learning rate over $\{10^{-i}\}_{i=1}^3$, and the learning rate decay is set to 0.999977. For \model{} and \linearmodel{}, we have an extra hyper-parameter $N$, which is searched over $\{2, 5, 10\}$. We fixed the total number of iterations $T$ to 5000. The experiments are executed for 10 times with different random seeds and the averaged results are reported in this paper.

\subsubsection{Evaluations. } To evaluate an OL2R model, we report both the offline and online performance during the interactions. The offline performance is evaluated in an ``online'' fashion where the newly updated ranker is evaluated on a hold-out testing set against its ground-truth relevance labels. This measures how fast an OL2R model improves its ranking quality. Such a setting can be viewed as using one portion of the traffic for model update, while serving another portion with the latest model. NDCG@10 is used to assess the ranking performance. In addition to the offline evaluation, we also evaluate the models' online result serving quality. This reflects user experience during the interactions and thus should be seriously considered. Sacrificing user experience for model training will compromise the goal of OL2R.  We adopt the cumulative Normalized Discounted Cumulative Gain to assess models' online performance. For $T$ rounds, the cumulative NDCG is calculated as
\begin{equation*}
    \text{Cumulative NDCG} = \sum\nolimits_{t=1}^T \text{NDCG}(\tau_t) \cdot \gamma^{(t-1)},
\end{equation*}
which computes the expected utility a user receives with a probability $\gamma$ that he/she stops searching after each query~\citep{oosterhuis2018differentiable}. Following the previous work~\citep{oosterhuis2018differentiable, wang2019variance, wang2018efficient, jia2021pairrank, jia2022neural}, we set $\gamma = 0.9995$.

\subsection{Experiment Results}

\begin{figure*}[t]
	\centering
		\subfigure{\includegraphics[width=\linewidth]{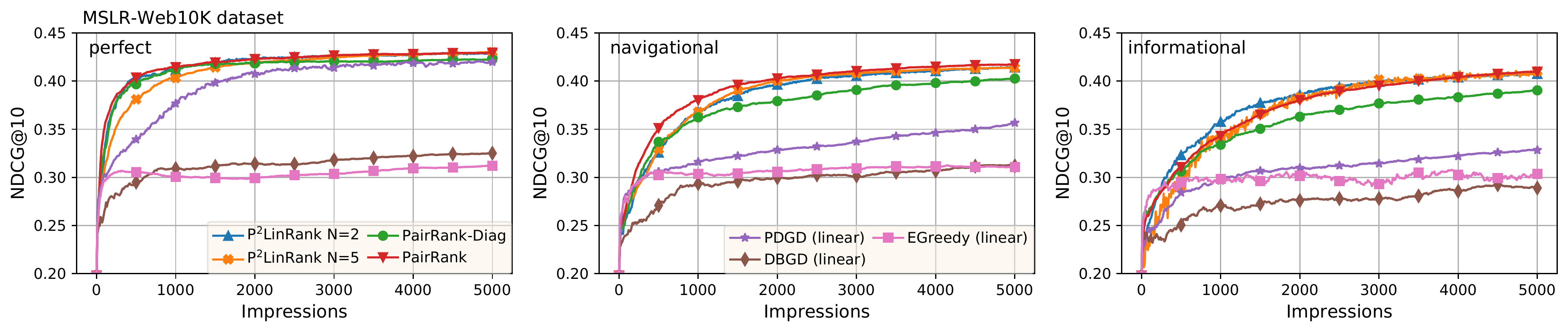}
		\label{fig:offline_web10k_linear}}\\ 
		\subfigure{\includegraphics[width=\linewidth]{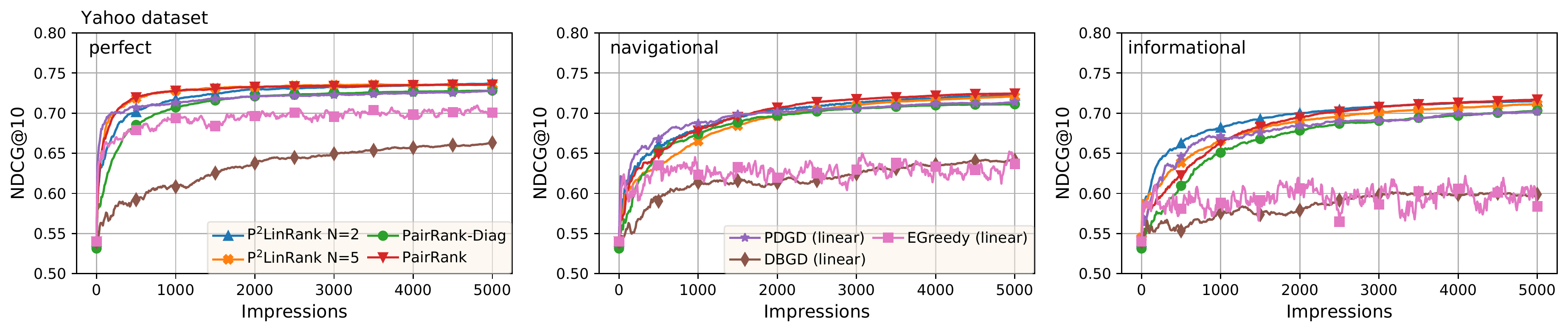}
		\label{fig:offline_yahoo_linear}}
	\caption{Offline performance of linear OL2R models on Web10K and Yahoo benchmark datasets.}\label{fig:offline_linear} 

		\subfigure{\includegraphics[width=\linewidth]{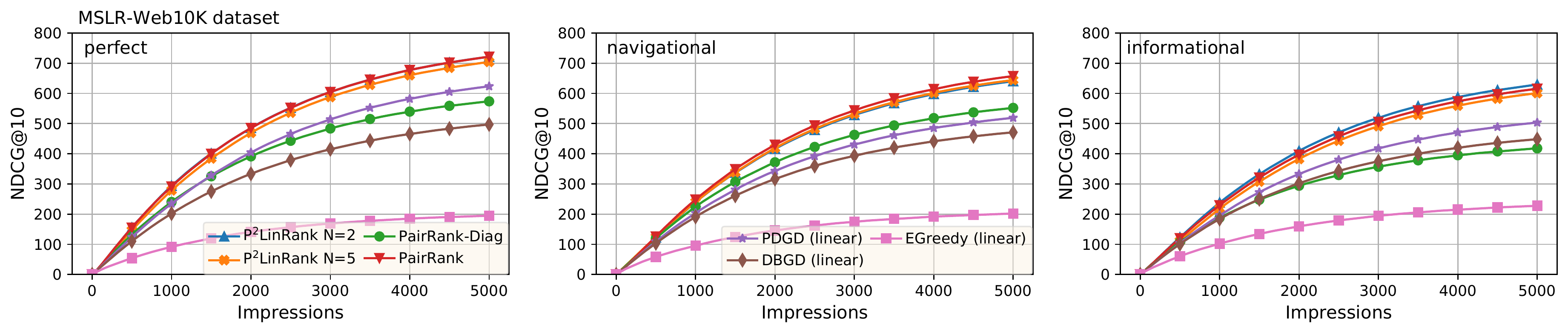}
		\label{fig:online_web10k_linear}}\\
		\subfigure{\includegraphics[width=\linewidth]{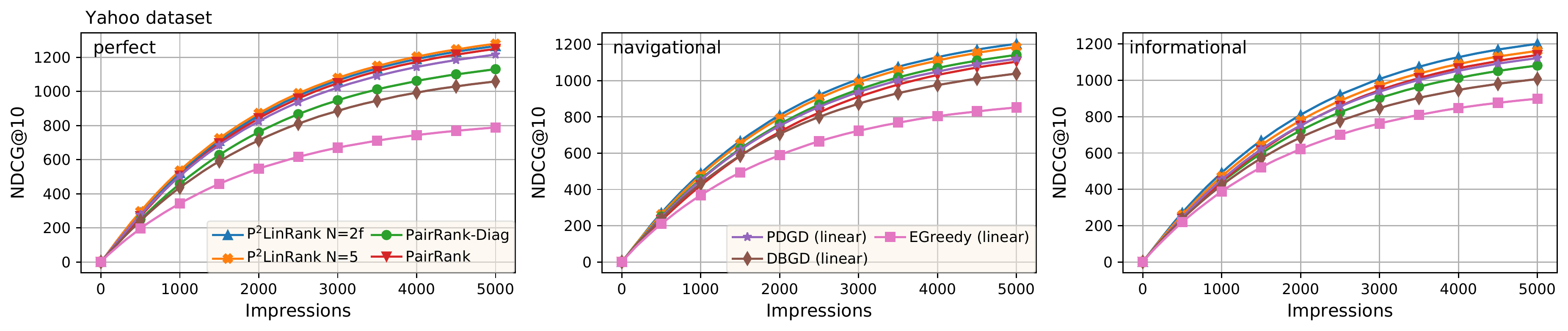}
		\label{fig:online_yahoo_linear}}
	\caption{Online performance of linear OL2R models on Web10K and Yahoo benchmark datasets.}\label{fig:online_linear} 
\end{figure*}

\subsubsection{Offline and online performance}
We first compare our proposed model with the baselines using a linear ranker. The results are reported in Figure~\ref{fig:offline_linear} and \ref{fig:online_linear}. For \linearmodel{}, we reported the best performance with $N=2$ and $N=5$. We can clearly observe \linearmodel{} maintained PairRank's strong advantage over other OL2R solutions, including $\epsilon$-Greedy, DBGD, and PDGD, in both online and offline evaluations across three click models. It is also obvious that a straightforward perturbation of model's output, i.e., $\epsilon$-Greedy, basically led to the worst OL2R performance, although it is often the default choice for exploration in online learning.

\begin{figure*}[t]
	\centering
		\subfigure{\includegraphics[width=\linewidth]{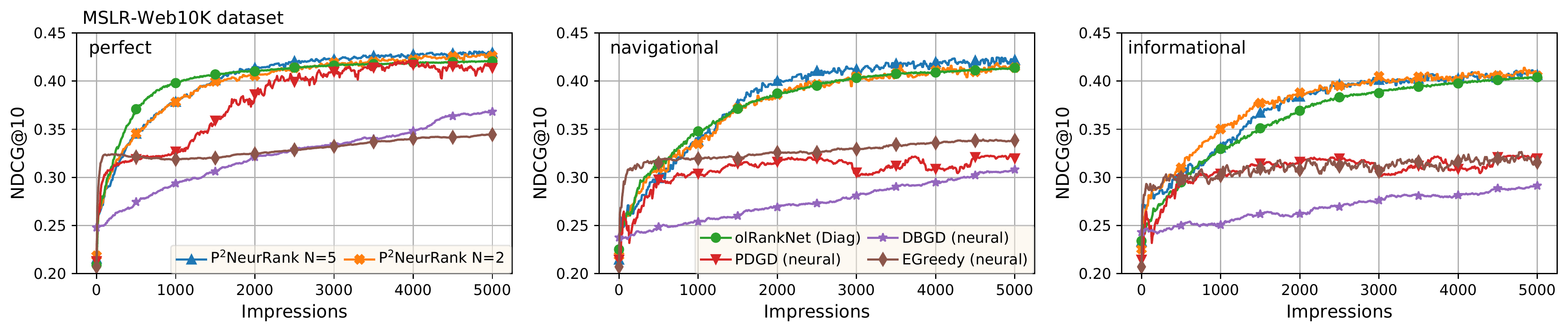}
		\label{fig:offline_web10k_neural}}\\  
		\subfigure{\includegraphics[width=\linewidth]{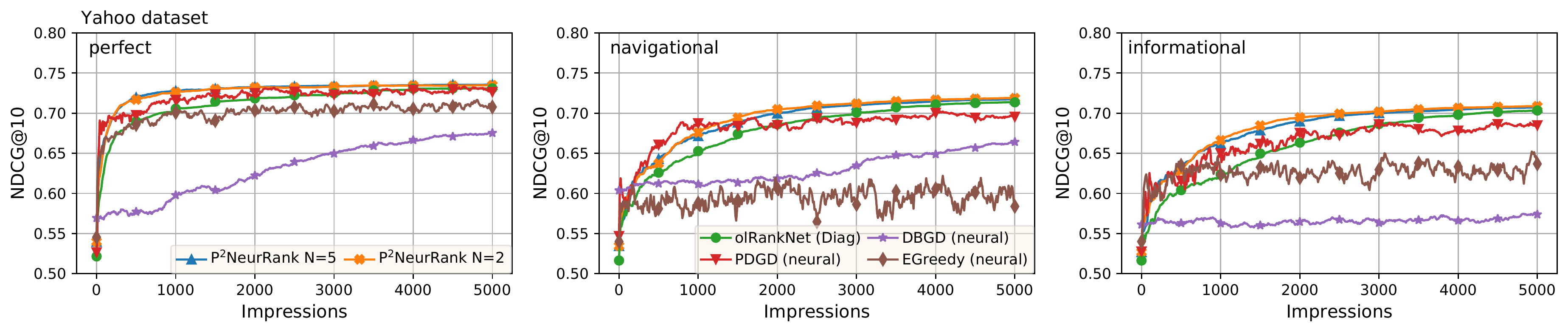}
		\label{fig:offline_yahoo_neural}}
	\caption{Offline performance of neural OL2R models on Web10K and Yahoo benchmark datasets.}\label{fig:offline_neural} 

		\subfigure{\includegraphics[width=\linewidth]{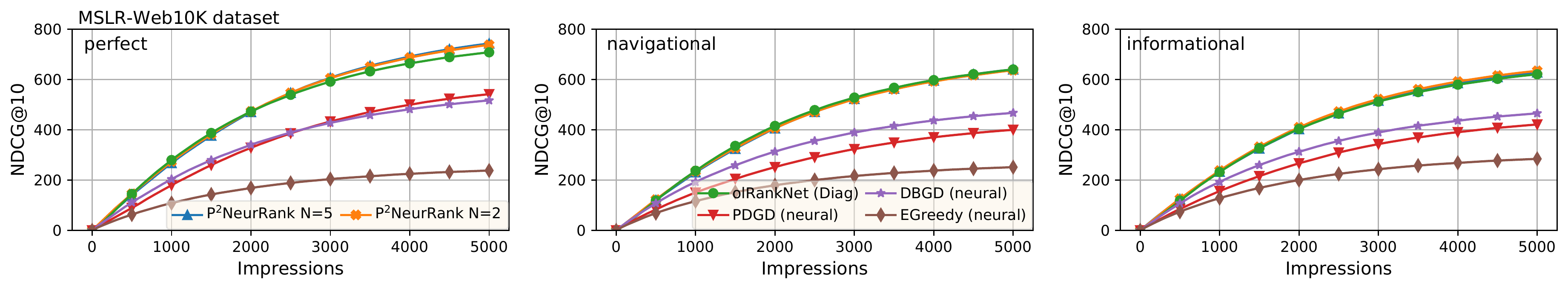}
		\label{fig:online_web10k_neural}}\\  
		\subfigure{\includegraphics[width=\linewidth]{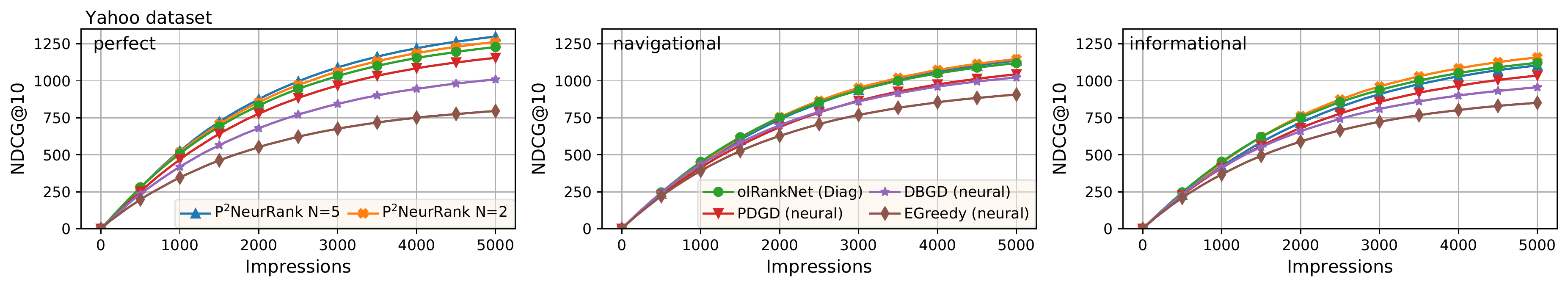}
		\label{fig:online_yahoo_neural}}
	\caption{Online performance of neural OL2R models on Web10K and Yahoo benchmark datasets.}\label{fig:online_neural} 
\end{figure*}

More importantly, PairRank with a diagonal approximated covariance matrix showed serious degradation in its ranking performance, especially its online performance. For example, on MSLR-Web10K dataset, PairRank with diagonal approximation was even worse than PDGD in the online evaluation under both perfect and informational click models. This means an approximated covariance matrix cannot accurately measure the ranker's estimation uncertainty. Furthermore, this inaccuracy's impact is not deterministic: under perfect click feedback, the seriously degenerated online performance together with mild decrease in offline performance suggest the model over explored; but under informational click feedback, both online and offline performance dropped, which suggests insufficient estimation. This demonstrates the complication of using approximations in OL2R solutions, which loses all theoretical guarantees in the original analysis. As a result, it also strongly suggests olRankNet might not be optimal in practice, given the diagonal approximation employed to make its computation feasible. This will be demonstrated in our experiments next.

It is worth noting that \linearmodel{} with $N=2$ already exhibits faster convergence than PairRank, and simply increasing $N$ does not necessarily further improve the model's performance. This result is very promising: as the only computational overhead in our perturbation-based exploration strategy is to estimate $N-1$ additional rankers, the actual added cost in practice is minimum when $N=2$. Similar observation is also obtained when applied to neural rankers.

\begin{figure}[t]
    \centering
    \includegraphics[width=\linewidth]{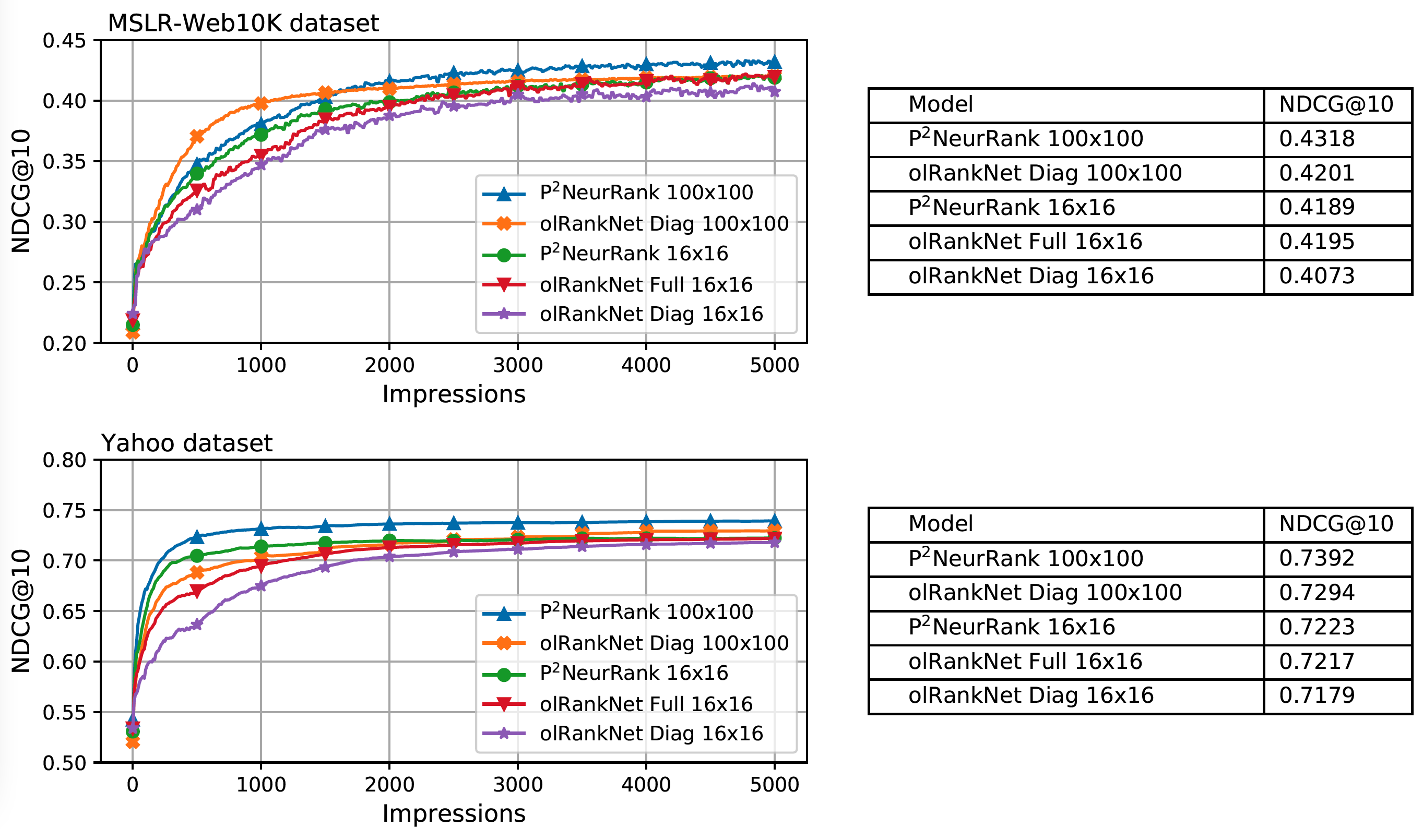}
    \caption{Performance comparison among different configurations of neural rankers.}
    \label{fig:neural_compare}
\end{figure}

In Figure~\ref{fig:offline_neural} and \ref{fig:online_neural}, we report the results obtained on the neural rankers. First of all, olRankNet and \model{} still showed significant improvement over other OL2R solutions, including $\epsilon$-Greedy, DBGD, MGD and PDGD. This means the pairwise exploration implemented in olRankNet is still effective for neural OL2R.
The most important finding in this experiment is that \model{} outperformed olRankNet, when the network width $m$ is set to 100. As we have repeatedly mentioned, though enjoying nice theoretical advantages, in practice it is impossible to use the required full covariance matrix to compute the confidence interval in olRankNet; as a result, the diagonal approximation creates an unknown gap from its theoretical guarantee to practical performance. 
In Figure~\ref{fig:neural_compare}, we compare the offline performance of \model{} with $m=100$, olRankNet with $m=100$, and neural models with simpler neural structures under the perfect click model. The results demonstrate that for the models with $16\times16$ neural network, the diagonal approximation hurts the performance compared to using the full covariance matrix. As proved in our theoretical analysis, \model{} enjoys the same theoretical regret guarantee as olRankNet; but because it does not need to endure the approximation, all its nice theoretical properties are preserved in its actual implementation. Compared to the results obtained in linear models, we have good reasons to believe olRankNet with full covariance matrix could perform even better, but with a much larger (if not infeasible) computational overhead.

Again the impact from the number of parallel rankers one needs to maintain for perturbation-based exploration in \model{} is still not sensitive. As shown in both Figure~\ref{fig:offline_neural} and \ref{fig:online_neural}, $N=2$ gave us the most promising empirical performance, with the minimum added computational overhead. This is a strongly desired property for applying \model{} in practice.

\begin{figure}[t]
    \centering
    \includegraphics[width=\linewidth]{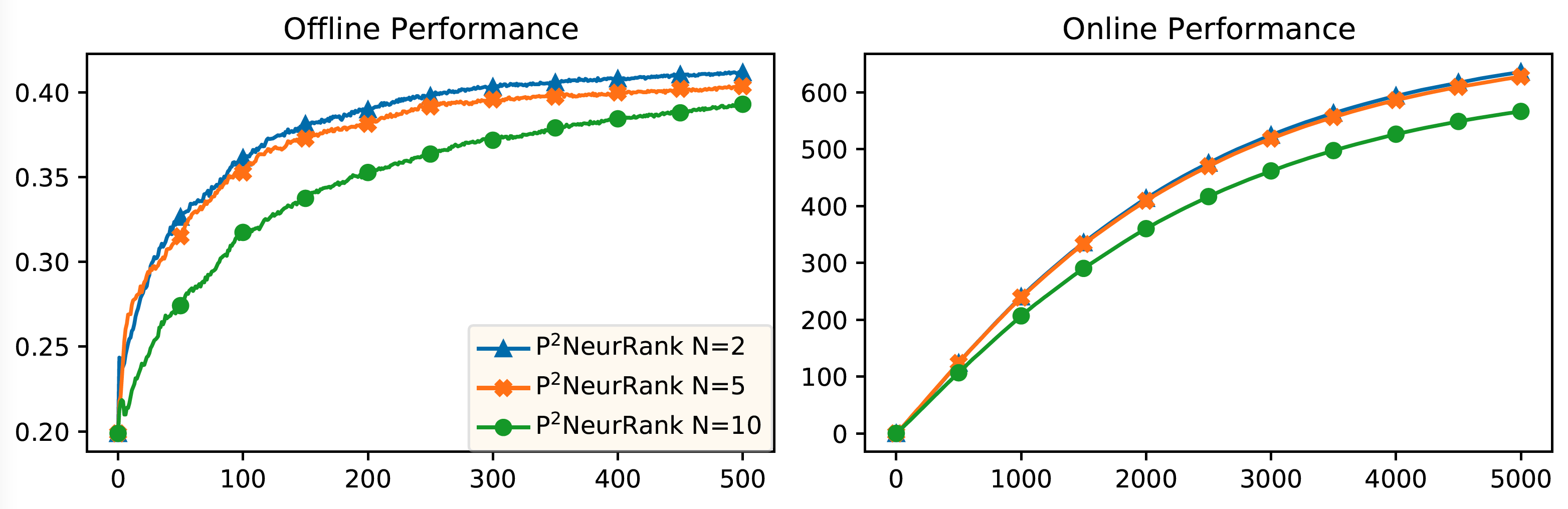}
    \caption{\model{} with different number of ensembled models under variance $\nu^2 = 0.1$.}
    \label{fig:variance_01}
    \includegraphics[width=\linewidth]{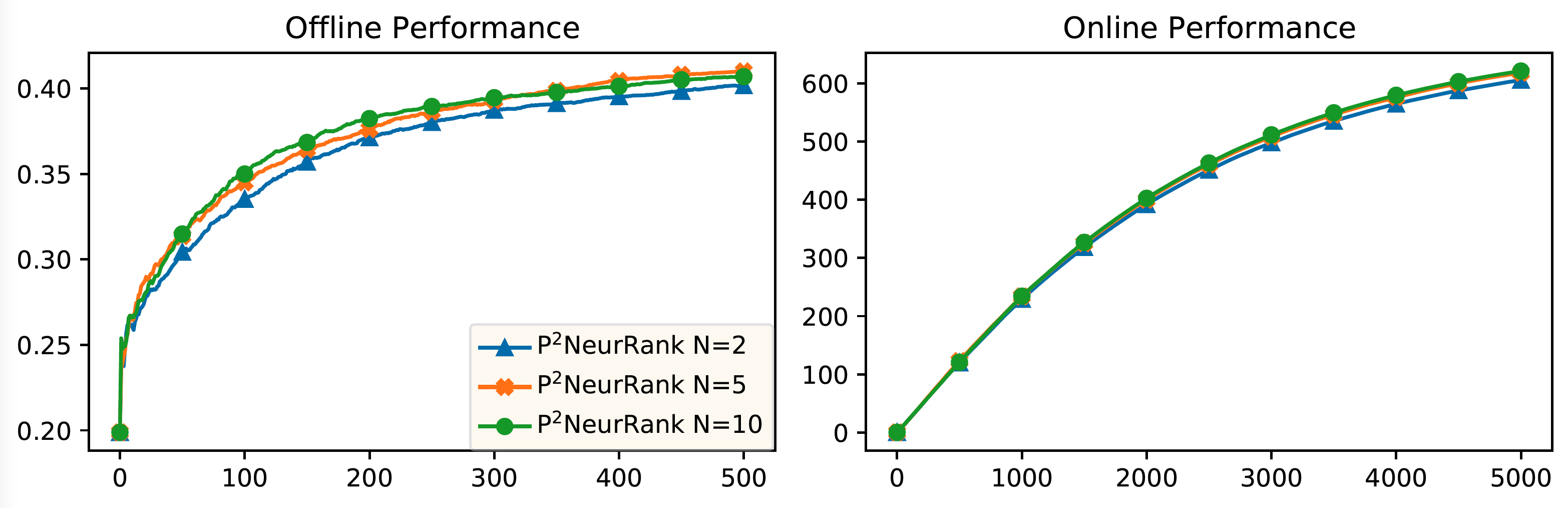}
    \caption{\model{} with different number of ensembled models under variance $\nu^2 = 0.01$.}
    \label{fig:variance_001}
\end{figure}

\subsubsection{Zoom into \model{}. }
In this experiment, we provide detailed analysis on \model{}. \model{} has only two hyper-parameters, in addition to those inherited from olRankNet, i.e., the number of parallel rankers $N$ and the scaled of pesudo noise $\nu^2$. 
In Figure~\ref{fig:variance_01} and \ref{fig:variance_001}, we report the online and offline performance of \model{} with varying value of $N$ for a fixed variance scale $\nu^2$ of the added noise. We can clearly observe that with a larger noise scale, e.g., $\nu = 0.1$, setting $N=2$ gives the best performance, comparing to $N=5$ and $N=10$. When the added noise scale is small, e.g., $\nu=0.01$, setting $N=10$ demonstrates better performance than those with fewer number of models. 
This indicates that the variance of the added noise $\nu$ and the number of parallel rankers $N$ together control the exploration in \model{}. A large variance scale, e.g., $\nu = 0.1$, together with too many models, e.g., $N=10$, lead to more aggressive exploration and less effective model training. A small variance, e.g., $\nu=0.01$, together with few models might not lead to sufficient exploration for model update, which also leads to worse performance. 
Therefore, in practice, the value of $\nu$ and $N$ should be carefully handled to perform effective exploration. And considering the added computational overhead, using fewer parallel rankers with larger scale of added noise should be a preferred solution. 

\begin{figure}[t]
    \centering
    \includegraphics[width=0.8\linewidth]{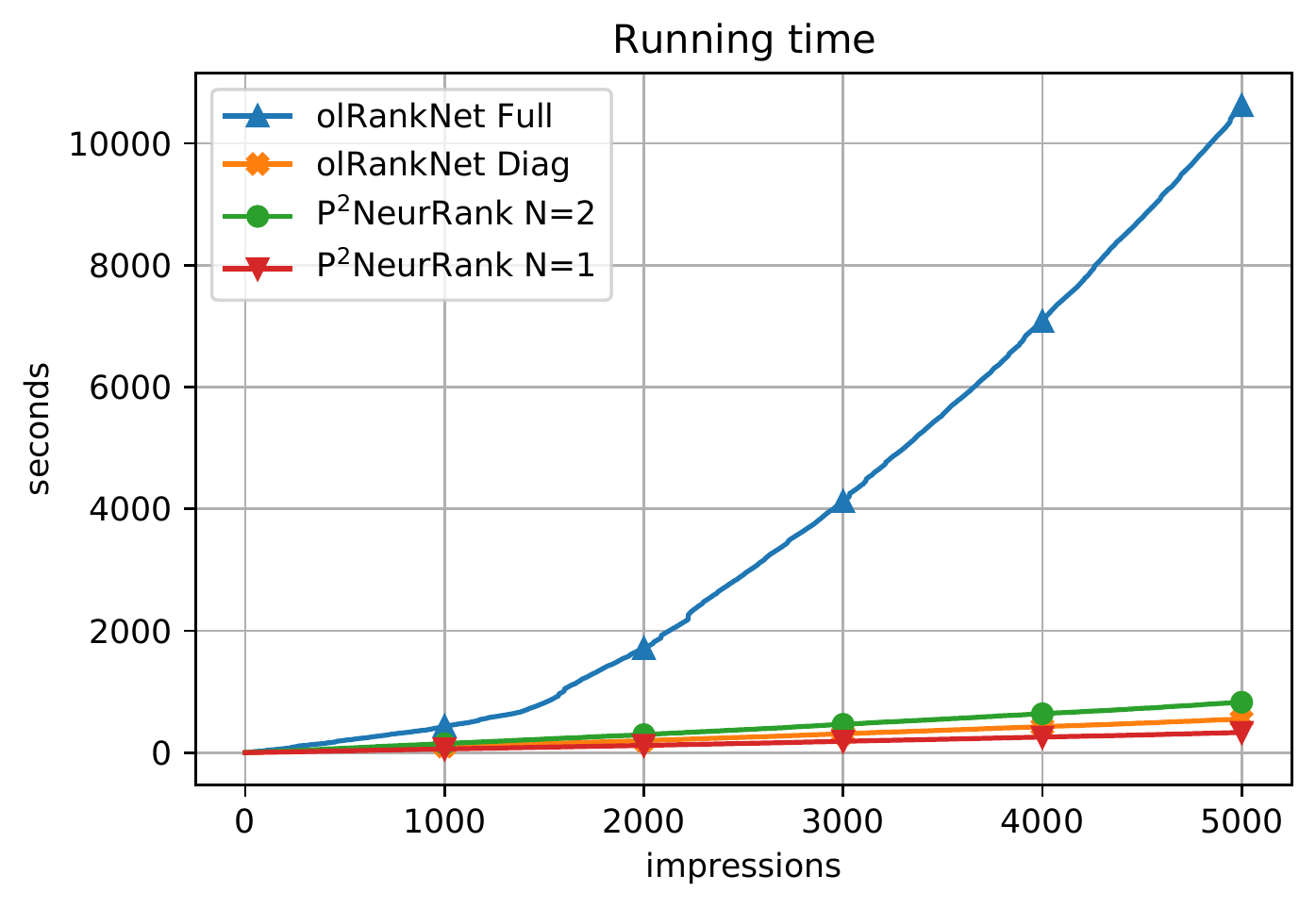}
    \caption{Efficiency comparison}
    \label{fig:efficiency}
\end{figure}

\subsubsection{Efficiency comparison.}
In this section, we compare the running time of our proposed \model{} and the olRankNet models. We performed the experiments on a NVIDIA GeForce RTX 2080Ti graphical card.
As discussed before, with complex neural networks, e.g., $m = 100$, it is impossible to perform the inverse of the full covariance matrix due to both high space and time complexity. Therefore, to compare running time, we perform the experiments on the MSLR-Web10K dataset using a simpler neural network with $m = 16$, where the perfect click model is adopted to generate the clicks. The result is reported in Figure~\ref{fig:efficiency}. We compare the olRankNet with a full covariance matrix, olRankNet with diagonal covariance matrix, \model{} with N=2 and \model{} with N = 1. For \model{}, no extra computation is required for  exploration. Therefore, the running time of \model{} with N = 1 can be viewed as the time used for the model training. We can clearly notice the big gap between the running time of olRankNet with full covariance matrix and \model{} with N = 1, which indicates the computational overhead caused by constructing the confidence interval  with the full covariance matrix. Using diagonal approximation greatly reduces the total running time. However, according to our previous discussion, there is no theoretical performance guarantee for such an approximation, and our empirical results show that the diagonal approximation often leads to decreased performance in both offline and online evaluations. On the other hand, \model{} with N=2 takes slightly more time than olRankNet with diagonal approximation, while the empirical performance is significantly better (shown in Figure~\ref{fig:offline_neural} and Figure~\ref{fig:online_neural}). Besides, in practice, the $N$ models can be trained in parallel, which will further reduce the running time. This demonstrate the feasibility and advantage of our proposed OL2R model in real applications.

To further investigate this difference, we examined neural rankers' performance under different neural network configurations (mainly the network width), with and without diagonal approximation of the covariance matrix, in Figure \ref{fig:neural_compare}. In particular, we tested a small enough network structure of $16\times 16$, on which we can maintain a full covariance matrix for olRankNet. On both datasets, a full covariance matrix for olRankNet with a $16\times 16$ network structure led to better performance than its counterpart using the diagonal approximation, while our \model{} maintained very similar performance with the full covariance version of olRankNet. And a larger network width indeed led to consistently better performance; however, the advantage is seriously hampered by the diagonal approximation employed in olRankNet, because of the compromise it has to make subject to the computational overhead. Again our \model{} did not suffer from this limitation and thus maintained its encouraging ranking performance.   

\section{Conclusion}
In this work, we developed a provable efficient exploration strategy for neural OL2R based on bootstrapping. Previous solutions for the purpose either do not have theoretical guarantees \citep{yue2009interactively,oosterhuis2018differentiable} or are too expensive to be exactly implemented in practice \citep{jia2022neural}. Our solution has an edge on both sides: it is proved to induce a sublinear upper regret bound counted over the number of mis-ordered pairs during online result serving, and its added computational overhead is feasible. Our extensive empirical evaluations demonstrate that our perturbation-based exploration unleashes the power of neural rankers in OL2R, with minimally added computational overhead (e.g., oftentimes only one additional ranker is needed to introduce the required exploration). And our perturbation-based exploration is general and can also be used in linear models when the input feature dimension is very large.

Our current theoretical analysis depends on gradient descent over the entire training set for model update in each round, which is still expensive and should be further optimized. We would like to investigate the possibility of more efficient model update, e.g., online stochastic gradient descent or continual learning, and the corresponding effect on model convergence and regret analysis. In addition, how to generalize our neural ranker architecture to more flexible choices, e.g., recurrent neural networks and Transformers, is another important direction of our future work.  

\section{Acknowledgements}
We want to thank the reviewers for their insightful comments. This work is based upon work supported by National Science Foundation under grant IIS-2128019, IIS-1718216 and IIS-1553568, and Google Faculty Research Award.

\bibliography{sample-base}

\begin{thebibliography}{60}
\providecommand{\natexlab}[1]{#1}
\providecommand{\url}[1]{\texttt{#1}}
\expandafter\ifx\csname urlstyle\endcsname\relax
  \providecommand{\doi}[1]{doi: #1}\else
  \providecommand{\doi}{doi: \begingroup \urlstyle{rm}\Url}\fi

\bibitem[Abbasi-Yadkori et~al.(2011)Abbasi-Yadkori, P{\'a}l, and
  Szepesv{\'a}ri]{abbasi2011improved}
Yasin Abbasi-Yadkori, D{\'a}vid P{\'a}l, and Csaba Szepesv{\'a}ri.
\newblock Improved algorithms for linear stochastic bandits.
\newblock In \emph{Advances in Neural Information Processing Systems}, pages
  2312--2320, 2011.

\bibitem[Agichtein et~al.(2006)Agichtein, Brill, and
  Dumais]{agichtein2006improving}
Eugene Agichtein, Eric Brill, and Susan Dumais.
\newblock Improving web search ranking by incorporating user behavior
  information.
\newblock In \emph{Proceedings of the 29th ACM SIGIR}, pages 19--26. ACM, 2006.

\bibitem[Arora et~al.(2019)Arora, Du, Hu, Li, Salakhutdinov, and
  Wang]{arora2019exact}
Sanjeev Arora, Simon~S Du, Wei Hu, Zhiyuan Li, Ruslan Salakhutdinov, and
  Ruosong Wang.
\newblock On exact computation with an infinitely wide neural net.
\newblock In \emph{Advances in Neural Information Processing Systems}, 2019.

\bibitem[Ash et~al.(2021)Ash, Zhang, Goel, Krishnamurthy, and
  Kakade]{ash2021anti}
Jordan~T Ash, Cyril Zhang, Surbhi Goel, Akshay Krishnamurthy, and Sham Kakade.
\newblock Anti-concentrated confidence bonuses for scalable exploration.
\newblock \emph{arXiv preprint arXiv:2110.11202}, 2021.

\bibitem[Auer(2002)]{auer2002using}
Peter Auer.
\newblock Using confidence bounds for exploitation-exploration trade-offs.
\newblock \emph{Journal of Machine Learning Research}, 3\penalty0
  (Nov):\penalty0 397--422, 2002.

\bibitem[Burges(2010)]{burges2010ranknet}
Christopher~JC Burges.
\newblock From ranknet to lambdarank to lambdamart: An overview.
\newblock \emph{Learning}, 11\penalty0 (23-581):\penalty0 81, 2010.

\bibitem[Cao and Gu(2019)]{cao2019generalization2}
Yuan Cao and Quanquan Gu.
\newblock Generalization bounds of stochastic gradient descent for wide and
  deep neural networks.
\newblock In \emph{Advances in Neural Information Processing Systems}, 2019.

\bibitem[Cao and Gu(2020)]{cao2019generalization1}
Yuan Cao and Quanquan Gu.
\newblock Generalization error bounds of gradient descent for learning
  over-parameterized deep relu networks.
\newblock In \emph{the Thirty-Fourth AAAI Conference on Artificial
  Intelligence}, 2020.

\bibitem[Chapelle and Chang(2011)]{chapelle2011yahoo}
Olivier Chapelle and Yi~Chang.
\newblock Yahoo! learning to rank challenge overview.
\newblock In \emph{Proceedings of the Learning to Rank Challenge}, pages 1--24,
  2011.

\bibitem[Chapelle et~al.(2012)Chapelle, Joachims, Radlinski, and
  Yue]{chapelle2012large}
Olivier Chapelle, Thorsten Joachims, Filip Radlinski, and Yisong Yue.
\newblock Large-scale validation and analysis of interleaved search evaluation.
\newblock \emph{ACM TOIS}, 30\penalty0 (1):\penalty0 6, 2012.

\bibitem[Chen et~al.(2019)Chen, Cao, Zou, and Gu]{chen2019much}
Zixiang Chen, Yuan Cao, Difan Zou, and Quanquan Gu.
\newblock How much over-parameterization is sufficient to learn deep relu
  networks?
\newblock \emph{arXiv preprint arXiv:1911.12360}, 2019.

\bibitem[Daniely(2017)]{daniely2017sgd}
Amit Daniely.
\newblock {SGD} learns the conjugate kernel class of the network.
\newblock In \emph{Advances in Neural Information Processing Systems}, pages
  2422--2430, 2017.

\bibitem[Dehghani et~al.(2017)Dehghani, Zamani, Severyn, Kamps, and
  Croft]{dehghani2017neural}
Mostafa Dehghani, Hamed Zamani, Aliaksei Severyn, Jaap Kamps, and W~Bruce
  Croft.
\newblock Neural ranking models with weak supervision.
\newblock In \emph{Proceedings of the 40th International ACM SIGIR Conference
  on Research and Development in Information Retrieval}, pages 65--74, 2017.

\bibitem[Du et~al.(2019)Du, Lee, Li, Wang, and Zhai]{du2018gradientdeep}
Simon Du, Jason Lee, Haochuan Li, Liwei Wang, and Xiyu Zhai.
\newblock Gradient descent finds global minima of deep neural networks.
\newblock In \emph{International Conference on Machine Learning}, pages
  1675--1685, 2019.

\bibitem[Goodfellow et~al.(2016)Goodfellow, Bengio, and
  Courville]{goodfellow2016deep}
Ian Goodfellow, Yoshua Bengio, and Aaron Courville.
\newblock \emph{Deep Learning}.
\newblock MIT Press, 2016.
\newblock \url{http://www.deeplearningbook.org}.

\bibitem[Guo et~al.(2009{\natexlab{a}})Guo, Liu, Kannan, Minka, Taylor, Wang,
  and Faloutsos]{guo2009click}
Fan Guo, Chao Liu, Anitha Kannan, Tom Minka, Michael Taylor, Yi-Min Wang, and
  Christos Faloutsos.
\newblock Click chain model in web search.
\newblock In \emph{Proceedings of the 18th WWW}, pages 11--20,
  2009{\natexlab{a}}.

\bibitem[Guo et~al.(2009{\natexlab{b}})Guo, Liu, and Wang]{guo2009efficient}
Fan Guo, Chao Liu, and Yi~Min Wang.
\newblock Efficient multiple-click models in web search.
\newblock In \emph{Proceedings of the 2nd WSDM}, pages 124--131,
  2009{\natexlab{b}}.

\bibitem[Guo et~al.(2020)Guo, Fan, Pang, Yang, Ai, Zamani, Wu, Croft, and
  Cheng]{guo2020deep}
Jiafeng Guo, Yixing Fan, Liang Pang, Liu Yang, Qingyao Ai, Hamed Zamani, Chen
  Wu, W~Bruce Croft, and Xueqi Cheng.
\newblock A deep look into neural ranking models for information retrieval.
\newblock \emph{Information Processing \& Management}, 57\penalty0
  (6):\penalty0 102067, 2020.

\bibitem[Hanin and Sellke(2017)]{hanin2017approximating}
Boris Hanin and Mark Sellke.
\newblock Approximating continuous functions by {ReLU} nets of minimal width.
\newblock \emph{arXiv preprint arXiv:1710.11278}, 2017.

\bibitem[Hofmann et~al.(2013)Hofmann, Whiteson, and
  de~Rijke]{hofmann2013balancing}
Katja Hofmann, Shimon Whiteson, and Maarten de~Rijke.
\newblock Balancing exploration and exploitation in listwise and pairwise
  online learning to rank for information retrieval.
\newblock \emph{Information Retrieval}, 16\penalty0 (1):\penalty0 63--90, 2013.

\bibitem[Huang et~al.(2013)Huang, He, Gao, Deng, Acero, and
  Heck]{huang2013learning}
Po-Sen Huang, Xiaodong He, Jianfeng Gao, Li~Deng, Alex Acero, and Larry Heck.
\newblock Learning deep structured semantic models for web search using
  clickthrough data.
\newblock In \emph{Proceedings of the 22nd ACM international conference on
  Information \& Knowledge Management}, pages 2333--2338, 2013.

\bibitem[Ishfaq et~al.(2021)Ishfaq, Cui, Nguyen, Ayoub, Yang, Wang, Precup, and
  Yang]{ishfaq2021randomized}
Haque Ishfaq, Qiwen Cui, Viet Nguyen, Alex Ayoub, Zhuoran Yang, Zhaoran Wang,
  Doina Precup, and Lin~F Yang.
\newblock Randomized exploration for reinforcement learning with general value
  function approximation.
\newblock \emph{arXiv preprint arXiv:2106.07841}, 2021.

\bibitem[Jacot et~al.(2018)Jacot, Gabriel, and Hongler]{jacot2018neural}
Arthur Jacot, Franck Gabriel, and Cl{\'e}ment Hongler.
\newblock Neural tangent kernel: Convergence and generalization in neural
  networks.
\newblock In \emph{Advances in neural information processing systems}, pages
  8571--8580, 2018.

\bibitem[Jia and Wang(2022)]{jia2022neural}
Yiling Jia and Hongning Wang.
\newblock Learning neural ranking models online from implicit user feedback.
\newblock \emph{arXiv preprint arXiv:2201.06658}, 2022.

\bibitem[Jia et~al.(2021)Jia, Wang, Guo, and Wang]{jia2021pairrank}
Yiling Jia, Huazheng Wang, Stephen Guo, and Hongning Wang.
\newblock Pairrank: Online pairwise learning to rank by divide-and-conquer.
\newblock In \emph{Proceedings of the Web Conference 2021}, pages 146--157,
  2021.

\bibitem[Jia et~al.(2022)Jia, Zhang, Zhou, Gu, and Wang]{jia2022perturbation}
Yiling Jia, Weitong Zhang, Dongruo Zhou, Quanquan Gu, and Hongning Wang.
\newblock Learning contextual bandits through perturbed rewards.
\newblock \emph{arXiv preprint arXiv:2201.09910}, 2022.

\bibitem[Joachims et~al.(2005)Joachims, Granka, Pan, Hembrooke, and
  Gay]{joachims2005accurately}
Thorsten Joachims, Laura Granka, Bing Pan, Helene Hembrooke, and Geri Gay.
\newblock Accurately interpreting clickthrough data as implicit feedback.
\newblock In \emph{Proceedings of the 28th ACM SIGIR}, pages 154--161. ACM,
  2005.

\bibitem[Joachims et~al.(2007)Joachims, Granka, Pan, Hembrooke, Radlinski, and
  Gay]{joachims2007evaluating}
Thorsten Joachims, Laura Granka, Bing Pan, Helene Hembrooke, Filip Radlinski,
  and Geri Gay.
\newblock Evaluating the accuracy of implicit feedback from clicks and query
  reformulations in web search.
\newblock \emph{ACM TOIS}, 25\penalty0 (2):\penalty0 7, 2007.

\bibitem[Kveton et~al.(2015)Kveton, Wen, Ashkan, and
  Szepesvari]{kveton2015combinatorial}
Branislav Kveton, Zheng Wen, Azin Ashkan, and Csaba Szepesvari.
\newblock Combinatorial cascading bandits.
\newblock In \emph{NIPS}, pages 1450--1458, 2015.

\bibitem[Kveton et~al.(2018)Kveton, Li, Lattimore, Markov, de~Rijke,
  Szepesvari, and Zoghi]{kveton2018bubblerank}
Branislav Kveton, Chang Li, Tor Lattimore, Ilya Markov, Maarten de~Rijke, Csaba
  Szepesvari, and Masrour Zoghi.
\newblock Bubblerank: Safe online learning to rerank.
\newblock \emph{arXiv preprint arXiv:1806.05819}, 2018.

\bibitem[Kveton et~al.(2019{\natexlab{a}})Kveton, Szepesv\'{a}ri, Ghavamzadeh,
  and Boutilier]{kveton19perturbed}
Branislav Kveton, Csaba Szepesv\'{a}ri, Mohammad Ghavamzadeh, and Craig
  Boutilier.
\newblock Perturbed-history exploration in stochastic linear bandits.
\newblock In \emph{Proceedings of the 35th Conference on Uncertainty in
  Artificial Intelligence (UAI)}, page 176, 2019{\natexlab{a}}.

\bibitem[Kveton et~al.(2019{\natexlab{b}})Kveton, Szepesvari, Vaswani, Wen,
  Lattimore, and Ghavamzadeh]{kveton2019garbage}
Branislav Kveton, Csaba Szepesvari, Sharan Vaswani, Zheng Wen, Tor Lattimore,
  and Mohammad Ghavamzadeh.
\newblock Garbage in, reward out: Bootstrapping exploration in multi-armed
  bandits.
\newblock In \emph{International Conference on Machine Learning}, pages
  3601--3610. PMLR, 2019{\natexlab{b}}.

\bibitem[Kveton et~al.(2020)Kveton, Zaheer, Szepesvári, Li, Ghavamzadeh, and
  Boutilier]{kveton20randomized}
Branislav Kveton, Manzil Zaheer, Csaba Szepesvári, Lihong Li, Mohammad
  Ghavamzadeh, and Craig Boutilier.
\newblock Randomized exploration in generalized linear bandits.
\newblock In \emph{Proceedings of the 22nd International Conference on
  Artificial Intelligence and Statistics}, 2020.

\bibitem[Lattimore and Szepesv\'{a}ri(2019)]{lattimore19bandit}
Tor Lattimore and Csaba Szepesv\'{a}ri.
\newblock \emph{Bandit Algorithms}.
\newblock Cambridge University Press, 2019.
\newblock In press.

\bibitem[Lattimore et~al.(2018)Lattimore, Kveton, Li, and
  Szepesvari]{lattimore2018toprank}
Tor Lattimore, Branislav Kveton, Shuai Li, and Csaba Szepesvari.
\newblock Toprank: A practical algorithm for online stochastic ranking.
\newblock In \emph{NIPS}, pages 3945--3954, 2018.

\bibitem[LeCun et~al.(2015)LeCun, Bengio, and Hinton]{lecun2015deep}
Yann LeCun, Yoshua Bengio, and Geoffrey Hinton.
\newblock Deep learning.
\newblock \emph{nature}, 521\penalty0 (7553):\penalty0 436, 2015.

\bibitem[Li et~al.(2018)Li, Lattimore, and Szepesv{\'a}ri]{li2018online}
Shuai Li, Tor Lattimore, and Csaba Szepesv{\'a}ri.
\newblock Online learning to rank with features.
\newblock \emph{arXiv preprint arXiv:1810.02567}, 2018.

\bibitem[Liang and Srikant(2016)]{liang2016deep}
Shiyu Liang and R~Srikant.
\newblock Why deep neural networks for function approximation?
\newblock \emph{arXiv preprint arXiv:1610.04161}, 2016.

\bibitem[Liu(2011)]{liu2011learning}
Tie-Yan Liu.
\newblock Learning to rank for information retrieval.
\newblock 2011.

\bibitem[Lu and Kawaguchi(2017)]{lu2017depth}
Haihao Lu and Kenji Kawaguchi.
\newblock Depth creates no bad local minima.
\newblock \emph{arXiv preprint arXiv:1702.08580}, 2017.

\bibitem[Oosterhuis and de~Rijke(2018)]{oosterhuis2018differentiable}
Harrie Oosterhuis and Maarten de~Rijke.
\newblock Differentiable unbiased online learning to rank.
\newblock In \emph{Proceedings of the 27th ACM CIKM}, pages 1293--1302, 2018.

\bibitem[Pasumarthi et~al.(2019)Pasumarthi, Bruch, Wang, Li, Bendersky, Najork,
  Pfeifer, Golbandi, Anil, and Wolf]{pasumarthi2019tf}
Rama~Kumar Pasumarthi, Sebastian Bruch, Xuanhui Wang, Cheng Li, Michael
  Bendersky, Marc Najork, Jan Pfeifer, Nadav Golbandi, Rohan Anil, and Stephan
  Wolf.
\newblock Tf-ranking: Scalable tensorflow library for learning-to-rank.
\newblock In \emph{Proceedings of the 25th ACM SIGKDD International Conference
  on Knowledge Discovery \& Data Mining}, pages 2970--2978, 2019.

\bibitem[Qin and Liu(2013)]{qin2013introducing}
Tao Qin and Tie-Yan Liu.
\newblock Introducing letor 4.0 datasets, 2013.

\bibitem[Schuth et~al.(2014)Schuth, Sietsma, Whiteson, Lefortier, and
  de~Rijke]{schuth2014multileaved}
Anne Schuth, Floor Sietsma, Shimon Whiteson, Damien Lefortier, and Maarten
  de~Rijke.
\newblock Multileaved comparisons for fast online evaluation.
\newblock In \emph{Proceedings of the 23rd ACM CIKM}, pages 71--80. ACM, 2014.

\bibitem[Schuth et~al.(2016)Schuth, Oosterhuis, Whiteson, and
  de~Rijke]{schuth2016multileave}
Anne Schuth, Harrie Oosterhuis, Shimon Whiteson, and Maarten de~Rijke.
\newblock Multileave gradient descent for fast online learning to rank.
\newblock In \emph{Proceedings of the 9th ACM WSDM}, pages 457--466, 2016.

\bibitem[Severyn and Moschitti(2015)]{severyn2015learning}
Aliaksei Severyn and Alessandro Moschitti.
\newblock Learning to rank short text pairs with convolutional deep neural
  networks.
\newblock In \emph{Proceedings of the 38th international ACM SIGIR conference
  on research and development in information retrieval}, pages 373--382, 2015.

\bibitem[Telgarsky(2015)]{telgarsky2015representation}
Matus Telgarsky.
\newblock Representation benefits of deep feedforward networks.
\newblock \emph{arXiv preprint arXiv:1509.08101}, 2015.

\bibitem[Telgarsky(2016)]{telgarsky2016benefits}
Matus Telgarsky.
\newblock Benefits of depth in neural networks.
\newblock \emph{arXiv preprint arXiv:1602.04485}, 2016.

\bibitem[Wang et~al.(2018{\natexlab{a}})Wang, Langley, Kim, McCord-Snook, and
  Wang]{wang2018efficient}
Huazheng Wang, Ramsey Langley, Sonwoo Kim, Eric McCord-Snook, and Hongning
  Wang.
\newblock Efficient exploration of gradient space for online learning to rank.
\newblock In \emph{SIGIR 2018}, pages 145--154, 2018{\natexlab{a}}.

\bibitem[Wang et~al.(2019)Wang, Kim, McCord-Snook, Wu, and
  Wang]{wang2019variance}
Huazheng Wang, Sonwoo Kim, Eric McCord-Snook, Qingyun Wu, and Hongning Wang.
\newblock Variance reduction in gradient exploration for online learning to
  rank.
\newblock In \emph{SIGIR 2019}, pages 835--844, 2019.

\bibitem[Wang et~al.(2018{\natexlab{b}})Wang, Li, Golbandi, Bendersky, and
  Najork]{Wang2018Lambdaloss}
Xuanhui Wang, Cheng Li, Nadav Golbandi, Michael Bendersky, and Marc Najork.
\newblock The lambdaloss framework for ranking metric optimization.
\newblock In \emph{CIKM '18}, pages 1313--1322. ACM, 2018{\natexlab{b}}.

\bibitem[Williams(2012)]{williams2012multiplying}
Virginia~Vassilevska Williams.
\newblock Multiplying matrices faster than coppersmith-winograd.
\newblock In \emph{Proceedings of the forty-fourth annual ACM symposium on
  Theory of computing}, pages 887--898, 2012.

\bibitem[Yarotsky(2017)]{yarotsky2017error}
Dmitry Yarotsky.
\newblock Error bounds for approximations with deep {ReLU} networks.
\newblock \emph{Neural Networks}, 94:\penalty0 103--114, 2017.

\bibitem[Yarotsky(2018)]{yarotsky2018optimal}
Dmitry Yarotsky.
\newblock Optimal approximation of continuous functions by very deep {ReLU}
  networks.
\newblock \emph{arXiv preprint arXiv:1802.03620}, 2018.

\bibitem[Yue and Joachims(2009)]{yue2009interactively}
Yisong Yue and Thorsten Joachims.
\newblock Interactively optimizing information retrieval systems as a dueling
  bandits problem.
\newblock In \emph{ICML}, pages 1201--1208, 2009.

\bibitem[Zhang et~al.(2020)Zhang, Zhou, Li, and Gu]{zhang2020neural}
Weitong Zhang, Dongruo Zhou, Lihong Li, and Quanquan Gu.
\newblock Neural thompson sampling.
\newblock \emph{arXiv preprint arXiv:2010.00827}, 2020.

\bibitem[Zhou et~al.(2019)Zhou, Li, and Gu]{zhou2019neural}
Dongruo Zhou, Lihong Li, and Quanquan Gu.
\newblock Neural contextual bandits with ucb-based exploration.
\newblock \emph{arXiv preprint arXiv:1911.04462}, 2019.

\bibitem[Zhou et~al.(2020)Zhou, Li, and Gu]{zhou2020neural}
Dongruo Zhou, Lihong Li, and Quanquan Gu.
\newblock Neural contextual bandits with ucb-based exploration.
\newblock In \emph{International Conference on Machine Learning}, pages
  11492--11502. PMLR, 2020.

\bibitem[Zou and Gu(2019)]{zou2019improved}
Difan Zou and Quanquan Gu.
\newblock An improved analysis of training over-parameterized deep neural
  networks.
\newblock In \emph{Advances in Neural Information Processing Systems}, 2019.

\bibitem[Zou et~al.(2019)Zou, Cao, Zhou, and Gu]{zou2018stochastic}
Difan Zou, Yuan Cao, Dongruo Zhou, and Quanquan Gu.
\newblock Stochastic gradient descent optimizes over-parameterized deep {ReLU}
  networks.
\newblock \emph{Machine Learning}, 2019.

\end{thebibliography}

\appendix
\section{Appendix}

\subsection{Proof of Lemma~\ref{lemma:cb}}

\begin{proof}
According to the definition of $g_t(\btheta)$ in Section 4, we have the following equation for the auxiliary solution $\bar \btheta_t$, 
\begin{align*}
    g_t(\bar \btheta_t) = \sum_{s=-d+1}^{t-1}\sum_{(i, j) \in \Omega_s} \yijs \xijs
\end{align*}
Then, to bound the deviation caused by the observation noise, we have the following derivations for any input $\xb$
\begin{align*}
    &|\sigma(\xb^\top\bar\btheta_t) - \sigma(\xb^\top\btheta^*)|  \\
    \leq & k_{\mu} |\xb^\top(\bar \btheta_t - \btheta^*)| \\
    =& k_{\mu} |\xb^\top\Gb_t^{-1}(g_t(\bar \btheta_t) - g_t(\btheta^*))| \\
    \leq & \frac{k_{\mu}}{c_{\mu}}|\xb^\top\Ab_t^{-1}(g_t(\bar \btheta_t) - g_t(\btheta^*))| \\
    \leq & \frac{k_{\mu}}{c_{\mu}}\|\xb\|_{\Ab_t^{-1}}\|g_t(\bar \btheta_t) - g_t(\btheta^*)\|_{\Ab_t^{-1}} 
\end{align*}
The first inequality is due to the Lipschitz continuity of the logistic function. As logistic function is continuously differentiable, the second equality is according to the Fundamental Theorem of Calculus, where $\Gb_t = \sum_{s=1}^{t-1}\sum_{(i, j)\in \Omega_s} \dot\sigma({\xijs}^\top\btheta) {\xijs}{\xijs}^\top + \lambda\Ib$.
In the third inequality, $\Ab_t = \sum_{s=1}^{t-1}\sum_{(i, j)\in \Omega_s} {\xijs}{\xijs}^\top + \lambda\Ib$. And this inequality holds as $\Gb_t \succ c_{\mu} \Ab_t$, with  $c_{\mu} = \inf_{\btheta \in \bTheta} \dot{\sigma}(\bx^\top\btheta)$.

Next, we will bound $\|g_t(\bar \btheta_t) - g_t(\btheta^*)\|_{\Ab_t^{-1}} $:
\begin{align*}
    &\|g_t(\bar \btheta_t) - g_t(\btheta^*)\|_{\Ab_t^{-1}} = \|\sum_{s=-d+1}^{t-1} \yijs\xijs - \sigma(\xijs^\top\btheta^*)\xijs\|_{\Ab_t^{-1}} \\
    \leq & \|\sum_{s=-d+1}^{0}  \sigma(\xijs^\top\btheta^*)\xijs\|_{\Ab_t^{-1}} + \|\sum_{s=1}^{t-1}  \epsilon_{ij}^s\xijs\|_{\Ab_t^{-1}} \leq d + R\sqrt{\log(\frac{\det(\Ab_t)}{\delta^2\det(\lambda\Ib)})}
\end{align*}
The first equality is based on the definition of function $g_t$, and $\bar\btheta_t$. The last inequality is according to the self-normalized bound for martingales in~\citep{abbasi2011improved}. This completes the proof.
\end{proof}
\subsection{Proof of Lemma~\ref{lemma:uncertain}}

\begin{proof}
According to Lemma~\ref{lemma:cb}, with probability at least $1 - \delta$, event $E_t$ defined in Section 4 occurs. Under event $E_t$, for document pair $(i, j)$ satisfying $i \succ j$ for the given query (e.g., $\sigma(\xijt^\top\btheta^*) > \frac{1}{2}$), we first analyze the probability that at least one estimate $\sigma(\xijt^\top\hat\btheta_t^{(n)}) > \frac{1}{2}$ for $n \in [N]$, e.g., $\PP(\max_{n\in[N]} \sigma(\xijt^\top\hat\btheta_t^{(n)})) > \frac{1}{2}$. For simplicity, in the following analysis, we use $\hat{\sigma}_t^{(n)}$, $\bar{\sigma}_t$ and $\sigma^*$ to present $\sigma\Big({\xijt}^\top\hat{\btheta}_t^{(n)}\Big)$, $\sigma\Big({\xijt}^\top\bar{\btheta}_t\Big)$ and $\sigma\Big({\xijt}^\top\btheta^*\Big)$ respectively.
\begin{align*}
    \PP(\max_{n\in[N]} \hat{\sigma}_t^{(n)} \geq \frac{1}{2} ) = 1 - \prod_{n=1}^N \PP(\sigma_t^{(n)} < \frac{1}{2})
\end{align*}
For any $n \in [N]$, we have the following bound for $\PP(\sigma_t^{(n)} < \frac{1}{2})$ .
\begin{align*}
    \PP(\sigma_t^{(n)} < \frac{1}{2}) \leq& \PP(\sigma_t^{(n)} < \sigma^*) = \PP(\sigma_t^{(n)} - \bar\sigma_t< \sigma^* - \bar\sigma_t) \\
    \leq& \PP(\sigma_t^{(n)} - \bar\sigma_t< \alpha_t \|\xijt\|_{\Ab_t^{-1}})\\
    \leq & \PP(c_{\mu}\xijt^\top(\hat \btheta_t - \bar\btheta) < \alpha_t \|\xijt\|_{\Ab_t^{-1}}) \\
    \leq & \PP(c_{\mu}\xijt^\top\Gb_t^{-1}(g_t(\hat \btheta_t) - g_t(\bar\btheta)) < \alpha_t \|\xijt\|_{\Ab_t^{-1}})\\
    \leq & \PP(\frac{c_{\mu}}{k_{\mu}}\xijt^\top\Ab_t^{-1}(g_t(\hat \btheta_t) - g_t(\bar\btheta)) < \alpha_t \|\xijt\|_{\Ab_t^{-1}}) \\
    = & \PP(U(\xijt) < \frac{k_{\mu}}{c_{\mu}}\alpha_t \|\xijt\|_{\Ab_t^{-1}})
\end{align*}
where $U(\xijt) = \xijt^\top\Ab_t^{-1}(g_t(\hat \btheta_t) - g_t(\bar\btheta))$. According to the definition of $\bar \btheta_t$, we know that,
\begin{align*}
    U(\xijt) = \xijt^\top\Ab_t^{-1}\sum_{s=-d+1}^{t-1} \gamma_{ij}^s\xijs
\end{align*}
It is easy to obtain that $\EE[U(\xijt)] = 0$ with $\EE[\gamma] = 0$. And the variance of $U(\xijt)$ is,
\begin{align*}
    Var[U(\xijt)] =& \nu^2 \xijt^\top\Ab_t(\sum_{s=-d+1}^{t-1}\xijs\xijs^\top)\Ab_t\xijt  \\
    =& \nu^2 \xijt^\top\Ab_t^{-1}(\lambda\Ib + \sum_{s=1}^{t-1}\xijs\xijs^\top)\Ab_t^{-1}\xijt  \\
    =& \nu^2\|\xijt\|_{\Ab_t^{-1}}^2
\end{align*}
Therefore, we have $U(\xijt) \sim \cN(0, \nu^2\|\xijt\|_{\Ab_t^{-1}}^2)$, and the probability can be upper bounded as,
\begin{align*}
    \PP(U(\xijt) < \frac{k_{\mu}}{c_{\mu}}\alpha_t \|\xijt\|_{\Ab_t^{-1}}) =& 1 - \PP(U(\xijt) > \frac{k_{\mu}}{c_{\mu}}\alpha_t \|\xijt\|_{\Ab_t^{-1}}) \\
    \leq & 1 - \frac{\exp(-\beta^2)}{4\sqrt{\pi}\beta}
\end{align*}
where $\beta = \frac{k_\mu\alpha_t}{c_\mu\nu}$. By chaining all the inequalities, we have that with $N \geq \log{\delta}/\log(1 - \exp(-\beta^2)/(4\sqrt{\pi}\beta))$, with probability at least $1 - \delta$, $\PP(\max_{n\in[N]} \hat{\sigma}_t^{(n)} \geq \frac{1}{2} )$.

Therefore, under event $E_t$, with $N \geq \log{\delta}/\log(1 - \exp(-\beta^2)/(4\sqrt{\pi}\beta))$, based on the definition of $\omega_t^u$ in Section 3.2, we know that for document $i$ and $j$ at round $t$, $(i, j) \in \omega_t^c$ if and only if $\min_{n\in[N]} \sigma\Big(\xijt^\top\hat{\btheta}_t^{(n)}\Big) > \frac{1}{2}$. 
%For simplicity, in the following analysis, we use $\hat{\sigma}_t^{(n)}$, $\bar{\sigma}_t$and $\sigma^*$ to present $\sigma\Big({\xijt}^\top\hat{\btheta}_t^{(n)}\Big)$, $\sigma\Big({\xijt}^\top\bar{\btheta}_t\Big)$and $\sigma\Big({\xijt}^\top\btheta^*\Big)$ respectively. 
Then the probability of being uncertain can be bounded as,
\begin{align*}
    &\mathbb{P}\big((i, j) \in \omega_t^u\big) = 1 -  \mathbb{P}\big(\min_{n\in[N]}  \hat{\sigma}_t^{(n)} > 1/2\big)\\
    =& 1 -  \prod_{n=1}^N \mathbb{P}\big( \hat{\sigma}_t^{(n)} > 1/2\big) = 1 -  \big(\mathbb{P}\big( \hat{\sigma}_t^{(n)} > 1/2\big)\big)^N
\end{align*}
Based on the definition of $\Delta_{\min}$, $\sigma^*- 1/2 \geq \Delta_{\min}$ and $\sigma^*- \bar{\sigma}_t \leq CB$. And according to the random matrix theory and Lemma 2 in \citep{jia2021pairrank}, when $t > t^\prime$, $\Delta_{\min} - CB > 0$ which can be viewed as a constant $c\Delta_{\min}$. Therefore, we have the following inequalities,
\begin{align*}
    \mathbb{P}\big( \hat{\sigma}_t^{(m)} > 1/2\big) =& \mathbb{P}\big( \hat{\sigma}_t^{(m)} - \sigma^* + \sigma^*- \bar{\sigma}_t + \bar{\sigma}_t > 1/2\big) \\
    \geq & \mathbb{P}\big( \bar{\sigma}_t - \hat{\sigma}_t^{(m)} < \Delta_{\min} - CB\big) = 1 - \frac{1}{2}\mathbb{P}\big( |\bar{\sigma}_t - \hat{\sigma}_t^{(m)}| \geq \Delta_{\min} - CB\big)\\
    \geq & 1 - \frac{1}{2}\mathbb{P}\big( k_{\mu}|U(x)| \geq \Delta_{\min} - CB\big)\\
    \geq & 1 - \exp(-\frac{c_{\mu}^2c^2\Delta_{\min} ^2}{2k_{\mu}^2\nu^2\|x\|_{\bA_t^{-1}}^2})
\end{align*}

Let $B = \frac{c_{\mu}^2c^2\Delta_{\min}^2}{2k_{\mu}^2\nu^2\|x\|_{\bA_t^{-1}}^2}$, we have the probability of being uncertain rank order upper bounded as,
\begin{align*}
    \mathbb{P}\big((i, j) \in \omega_t^u\big) \leq& 1 -  \big(1 - \exp(-B))\big)^N \\
    = & 1 - \exp N\log(1 - \exp(-B)) \\
    \leq & -N\log(1 - \exp(-B)) \leq -N\log\exp(-1/B))
\end{align*}

Therefore, $\mathbb{P}\big((i, j) \in \omega_t^u\big) \leq \min \Big\{1, \frac{2Nk_{\mu}^2\nu^2\|x_{i,j}\|_{\bA_t^{-1}}^2}{c_{\mu}^2c^2\Delta_{\min}^2}\Big\}$. This completes the proof.
\end{proof}

\subsection{Proof of Theorem~\ref{theorem}}
\begin{proof}

Once the certain and uncertain rank orders are determined, our proposed model will generate the ranked list by topological sort with respect to the certain rank orders. Therefore, the regret only comes from the uncertain rank orders. In each round of result serving, as the model $\btheta_t$ would not change until the next round, the expected number of uncertain rank orders $U_t$,  can be estimated by summing the uncertain probabilities over all possible pairwise comparisons under the current query $q_t$, e.g., $\EE[U_t] =   \frac{V_t(V_t-1)}{2} \mathbb{P}((i, j) \in \omega_t^u)$.

Based on Lemma~\ref{lemma:uncertain}, the cumulative number of mis-ordered pairs can be bounded by the probability of observing uncertain rank orders in each round, which shrinks with more observations become available over time, 
\begin{align*}
    \EE\big[\sum\nolimits_{s=t^\prime}^{T} U_t\big] 
    \leq& \EE\big[\frac{1}{2}\sum\nolimits_{s=t^\prime}^{T} \sum\nolimits_{(i, j) \in [V_t]^2} \frac{2Nk_{\mu}^2\nu^2\|\xijs\|_{\bA_s^{-1}}^2}{c_{\mu}^2c^2\Delta_{\min}^2}\big].
\end{align*}

As $\Ab_t$ only contains information of observed document pairs so far, the number of mis-ordered pairs among the observed documents is guaranteed to be upper bounded. To reason about the number of mis-ordered pairs in those unobserved documents (i.e., from $o_t$ to $V_t$ for each query $q_t$), we leverage the constant $p^*$, which is defined as the minimal probability that all documents in a query are examined over time,
\begin{align*}
    &\EE\big[\sum\nolimits_{t=\tp}^T\sum\nolimits_{(i, j)\in[V_t]^2} \|\xijt\|_{\Ab_{t}^{-1}}\big] \\
    =& \EE\big[\sum\nolimits_{t=\tp}^T\sum\nolimits_{ (i, j)\in[V_t]^2} \|\xijt\|_{\Ab_{t}^{-1}} \times \EE\big[{p_t^{-1}}\textbf{1}\{o_t = V_t\}\big]\big] \\
    \leq & p^*{^{-1}}\EE\big[\sum\nolimits_{t=\tp}^T\sum\nolimits_{(i, j)\in[V_t]^2} \|\xijt\|_{\Ab_{t}^{-1}}\textbf{1}\{o_t = V_t\}\big]
\end{align*}

Besides, we only use the independent pairs $\Omega_t$ to update the model and the corresponding $\Ab_t$ matrix. Therefore, to bound the regret, the pairs can be divided into two parts based on whether they are belonging to the observed set $\Omega_t$. Then, we have the following inequalities,
\begin{align*}
    \sum\nolimits_{s=\tp}^T\sum\nolimits_{(i, j) \in \Omega_s}\sum\nolimits_{k\in [V_t] \setminus \{i, j\} } 2{\bx_{ik}^s}^\top\bA_s^{-1}\bx_{jk}^s 
    \leq  \sum\nolimits_{s=\tp}^T {(V_{\max}^2 - 2V_{\max})P^2 }/{\lambda_{\min}(\bA_s)}.
\end{align*}

And for the pairs belonging to the $\{\Omega_s\}_{s=1}^T$, based on Lemma 10 and Lemma 11 in \citep{abbasi2011improved}, we have,
\begin{align*}
    \sum\nolimits_{s=\tp}^T\sum\nolimits_{(i, j) \in \Omega_s} (V_t - 1)\Vert\xijs\Vert_{\bA_s^{-1}}^2 \leq 2dV_{\max}\log(1 + \frac{o_{\max}TP^2}{2d\lambda})
\end{align*}

Chaining all the inequalities, we have when event $E_t$ happens, the regret can be upper bounded as,
\begin{align*}
    R_T \leq & R^{\prime} + \sum\nolimits_{s=\tp}^T r_s \leq R^\prime + \frac{1}{p^*}2dV_{\max}C\log(1 + \frac{o_{\max}TP^2}{2d\lambda})
\end{align*}
where $C={2Nk_{\mu}^2\nu^2}/{c_{\mu}^2c^2\Delta_{\min}^2}$,  $R^{\prime} = \tp V_{\max}$, with $\tp$ defined in Lemma~\ref{lemma:uncertain}. This completes the proof.
\end{proof}

\subsection{Regret analysis of \model{}}
According to the setting of olRankNet~\citep{jia2022neural}, we assume for the learning to rank problem, there exists an unknown function $h(\cdot)$ that models the relevance quality of document $\xb$ under the given query $q$ as $h(\xb)$. In order to learn this function, we utilize a fully connected neural network $f(\xb;\btheta) = \sqrt{m}\Wb_L \phi(\Wb_{L-1} \phi(\dots \phi(\Wb_1\xb))$, where depth $L \geq 2$, $\phi(\xb) = \max\{\xb, 0\}$, and $\Wb_1 \in \RR^{m \times d}$, $\Wb_i \in \RR^{m \times m}$, $2\leq i \leq L-1$, $\Wb_L \in \RR^{m \times 1}$, and $\btheta = [\text{vec}(\Wb_1)^\top,\dots,\text{vec}(\Wb_L)^\top]^\top \in \RR^{p}$ with $p = m+md+m^2 (L-2)$. Without loss of generality, we assume the width of each hidden layer is the same as $m$, concerning the simplicity of theoretical analysis. We also denote the gradient of the neural network function as $\gb(\xb; \btheta) = \nabla_{\btheta} f(\xb; \btheta) \in \RR^p$.
Therefore, the loss function is, 
\begin{align}
     \cL^{(n)}_t(\btheta) =& \sum_{s=1}^{t-1}\sum_{(i, j) \in \Omega_s} -\big(\yijs + \gamma_{ij}^{s, (n)}\big)\log(\sigma(f_{ij}^s)) 
     - \big(1 - (\yijs + \gamma_{ij}^{s, (n)})\big)\log\big(1 - \sigma(f_{ij}^s)\big) \nonumber\\
     &+ \frac{m\lambda}{2}\sum_{l=1}^p (\eb_l^\top(\btheta-\btheta_0))^2,
\end{align}
where the $l_2$ regularization $\frac{m\lambda}{2}\|\btheta - \btheta_0\|^2$ in Eq~\eqref{eq:obj} is re-written as $\frac{m\lambda}{2}\sum_{l=1}^p (\eb_l^\top(\btheta-\btheta_0))^2$ with $\eb_l$ as the standard basis of the $p-$dimensional space. 

\begin{figure*}[t]
	\centering
		\subfigure{\includegraphics[width=\linewidth]{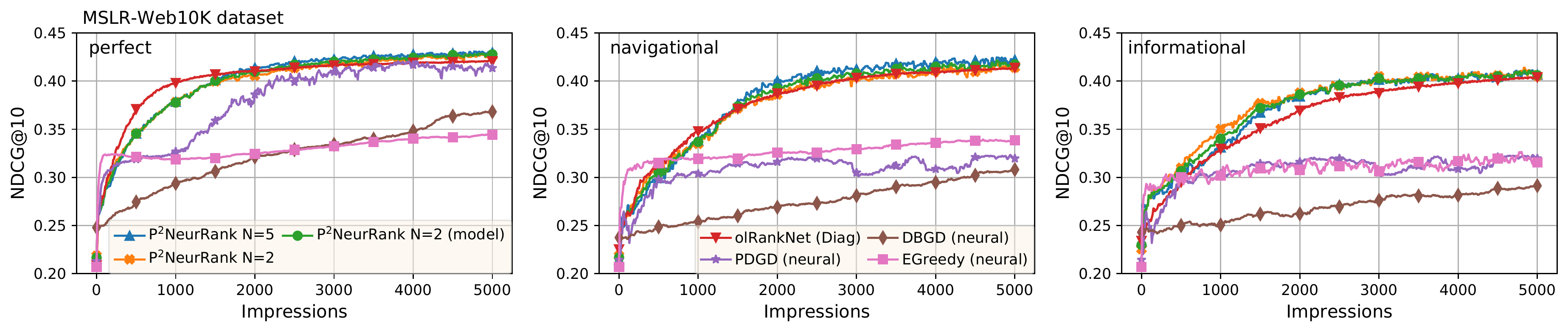}
		\label{fig:offline_web10k_neural_model}}\\  
		\subfigure{\includegraphics[width=\linewidth]{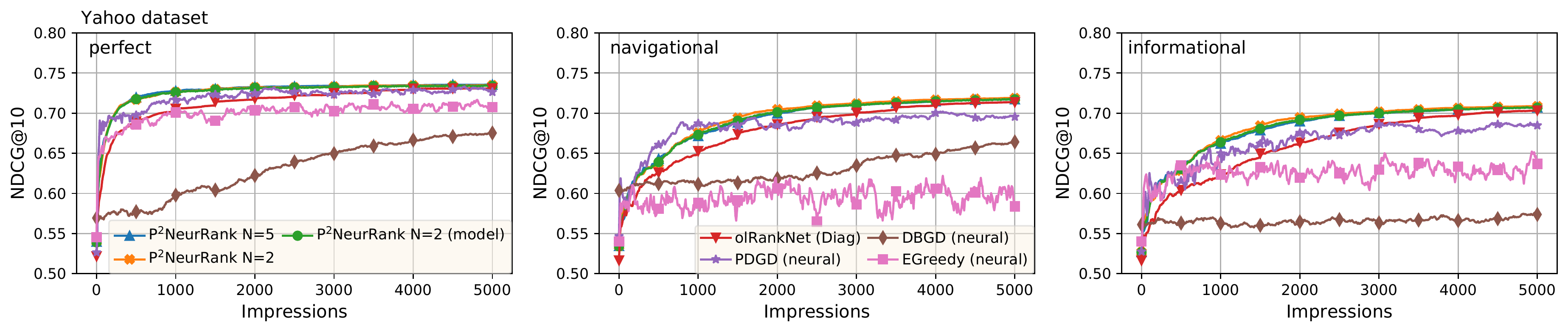}
		\label{fig:offline_yahoo_neural_model}}
	\caption{Offline performance of neural OL2R on two benchmark datasets.}\label{fig:offline_neural_model} 
	
		\subfigure{\includegraphics[width=\linewidth]{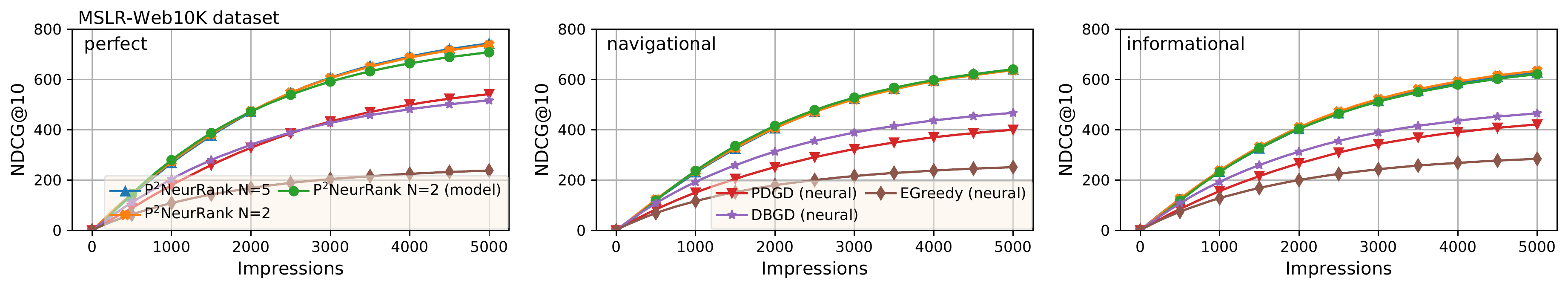}
		\label{fig:online_web10k_neural_model}}\\ 
		\subfigure{\includegraphics[width=\linewidth]{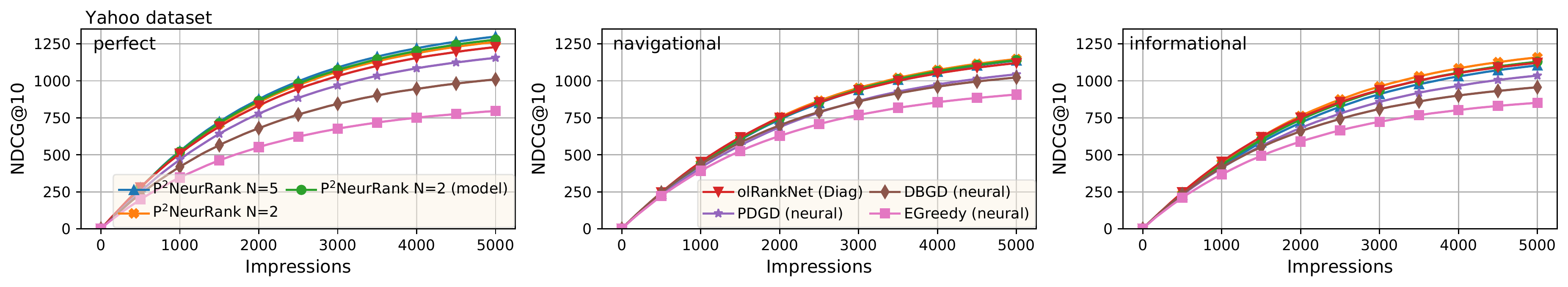}
		\label{fig:online_yahoo_neural_model}}
	\caption{Online performance of neural OL2R on two benchmark datasets.}\label{fig:online_neural_model} 
\end{figure*}

For the regret analysis of the neural ranker, \model{}, we slightly modified the perturbation strategy, where the loss function of the $n$-th model is,
\begin{align}
    \cL^{(n)}_t(\btheta) = &\sum_{s=1}^{t-1}\sum_{(i, j) \in \Omega_s} -\big(\yijs + \gamma_{ij}^{s, (n)}\big)\log(\sigma(f_{ij}^s)) 
     - \big(1 - (\yijs + \gamma_{ij}^{s, (n)})\big)\log\big(1 - \sigma(f_{ij}^s)\big) \nonumber \\
     &+ \frac{m\lambda}{2}\sum_{l=1}^p (\eb_l^\top(\btheta-\btheta_0) + 1/\sqrt{m\lambda}{\gamma}_l^{(n)}\eb_l)^2.
\end{align}
Different from the linear model where we only perturb the observed click feedback, we also perturbed the model by adding the pseudo noise to each dimension of the model parameter, $\tilde{\gamma}_l^{(n)}\eb_l$.  In particular, both the pseudo noise for the observed click feedback and the model parameters are sampled from the same distribution: $\gamma_{ij}^{s, (n)} \sim \cN(0, \nu^2)$ and ${\gamma}_l^{(n)} \sim \cN(0, \nu^2)$.  To first confirm this difference does not change the empirical performance of \model{}, we followed the same setting in Section \ref{sec:exp} to evaluate this new variant. Figure~\ref{fig:offline_neural_model} and Figure~\ref{fig:online_neural_model} show the comparison between the \model{} solution that only perturbs the click observations and that also perturbs the model parameters. We can observe that both solutions can achieve better offline and online performance than the existing baselines and the olRankNet model with diagonal approximation of the covariance matrix. For the two models with different choices of the perturbation, they have similar performance.

The following is the key conclusion of the regret analysis of \model{}. The difference between the analysis of \model{} and \linearmodel{} lies in the quantification of the probability that an estimated pairwise preference being uncertain. With a neural ranker, we need to take the neural network's approximation error into consideration when quantifying the probability. 

\begin{lemma}
\label{lemma:neuraluncertain}
There exist positive constants $c$, $\{C_i\}_{i=1}^4$, if the step size of gradient descent $\eta \leq C_1(TmL + m\lambda)^{-1}$ and 
$m \geq C_2\max\big\{ \lambda^{-1/2}L^{-3/2}(\log(TV_{\max}L^2/\delta_1))^{3/2}, T^7\lambda^{-7}L^{21}(\log m)^3\big\}$,
with $t^{\prime\prime} = \frac{2k_{\mu}P}{c_{\mu}\Delta_{\min}}\Big(\sqrt{R^2\log(1/\delta)} + \sqrt{\lambda}Q\Big) +  \Big( \frac{C_3\sqrt{d} + C_4\sqrt{\log(1/\delta)} + (P^2Rk_{\mu})/(\sqrt{\lambda}c_{\mu}\Delta_{\min})}{\lambda_{\min}(\Sigma)}\Big)^2$, $\delta \in (0, 1)$, for round $t \geq t^{\prime\prime}$, with probability at least $1 - \delta$, event $E_t$ holds with $\alpha_t$ defined in Lemma~\ref{lemma:cb}, with $N \geq \log{\delta}/\log(1 - \exp(-\beta^2)/(4\sqrt{\pi}\beta))$, where $\beta = \frac{k_\mu^2\alpha_t^2}{c_{\mu}^2\nu^2}$, for a document pair $(i, j)$ that $i \succ j$ for the given query, the probability that the estimated pairwise preference is uncertain is upper bounded as $\mathbb{P}\big((i, j) \in \omega_t^u\big) \leq \min \Big\{1, \frac{2N\nu^2\|\gb_{ij}^{t, 0}\|_{\bA_t^{-1}}^2}{c_{\mu}^2c^2\Delta_{\min}^2}\Big\}$,
where $\Delta_{\min} = \min\limits_{t\in T, (i, j) \in [V_t]^2}| \sigma(h_{ij}) - \frac{1}{2}|$ representing the smallest gap of pairwise difference between any pair of documents associated to the same query over time (across all queries), $\epsilon(m) = \bar C_1\Big(T^{7/6}m^{-1/6}\lambda^{-7/6}L^4\sqrt{\log(m)}(1 + \sqrt{T/\lambda}) + (1 - \eta m \lambda)^{J/2}\sqrt{TL/\lambda} + T^{1/6}m^{-1/6}\lambda^{-1/6}L^{7/2}\sqrt{\log(m)}S + T^{2/3}m^{-1/6}\lambda^{-2/3}L^3\sqrt{\log (m)}  \Big)$ is the approximation error of the neural network.
\end{lemma}

\subsubsection{Proof of Lemma~\ref{lemma:neuraluncertain}}

To quantify the probability that an estimated pairwise preference being uncertain, we start from introducing the background of neural tangent kernel technique.

We first assume that there are $n_T$ possible documents to be evaluated during the model learning. It is easy to conclude that $n_T = \sum_{t=1}^T V_T \leq TV_{\max}$. Now, we introduce the neural tangent kernel matrix defined on the $n_T$ possible query-document feature vectors across $T$ rounds, $\{\xb_i\}_{i=1}^{n_T}$.

\begin{definition}[\citet{jacot2018neural,cao2019generalization2}]\label{def:ntk}
Let $\{\xb_i\}_{i=1}^{n_T}$ be the set of all pairwise document feature vectors.  Define
\begin{align*}
    \tilde \Hb_{i,j}^{(1)} &= \bSigma_{i,j}^{(1)} = \la \xb_i, \xb_j\ra, ~~~~~~~~  \Bb_{i,j}^{(l)} = 
    \begin{pmatrix}
    \bSigma_{i,i}^{(l)} & \bSigma_{i,j}^{(l)} \\
    \bSigma_{i,j}^{(l)} & \bSigma_{j,j}^{(l)} 
    \end{pmatrix},\notag \\
    \bSigma_{i,j}^{(l+1)} &= 2\EE_{(u, v)\sim N(\zero, \Bb_{i,j}^{(l)})} \left[\phi(u)\phi(v)\right],\notag \\
    \tilde \Hb_{i,j}^{(l+1)} &= 2\tilde \Hb_{i,j}^{(l)}\EE_{(u, v)\sim N(\zero, \Bb_{i,j}^{(l)})} \left[\dot \phi(u)\dot \phi(v)\right] + \bSigma_{i,j}^{(l+1)}.\notag
\end{align*}
Then, $\Hb = (\tilde \Hb^{(L)} + \bSigma^{(L)})/2$ is called the \emph{neural tangent kernel (NTK)} matrix on the context set $\{x_i\}_{i=1}^{n_T}$. 
\end{definition}

We also need the following assumption on the NTK matrix and the corresponding feature set.

\begin{assumption}\label{assumption:input}
$\Hb \succeq \lambda_0\Ib$; 
moreover, for any $1 \leq i \leq n$, $\|\xb_i\|_2 = 1$ and $[\xb_i]_j =[\xb_i]_{j+d/2}$.
\end{assumption}

With this assumption, the NTK matrix is assumed to be non-singular~\citep{du2018gradientdeep,arora2019exact, cao2019generalization2}. This assumption can be easily satisfied when no two query-document feature vectors are in parallel.
The second assumption is for convenience in analysis and can be easily satisfied by: for any context $\xb, \|\xb\|_2 = 1$, we can construct a new context $\xb' = [\xb^\top, \xb^\top]^\top/\sqrt{2}$. Equipped with this assumption, it can be verified that with $\btheta_0$ initialized as in Algorithm~\ref{alg:model}, $f(\xb_i; \btheta_0) = 0$ for any $i \in [n_T]$.

Besides, we need the following technical lemmas for the regret analysis.

\begin{lemma}[Lemma 5.1, \citet{zhou2020neural}]\label{lemma:equal}
There exists a positive constant $\bar C$ such that for any $\delta \in (0,1)$, if $m \geq \bar Cn_T^4L^6\log(n_T^2L/\delta)/\lambda_0^4$, then with probability at least $1-\delta$, there exists a $\btheta^* \in \RR^p$ such that for any $i \in [n_T]$, with $\hb = (h(\xb_1), \dots, h(\xb_n))$. 
\begin{align}
    h(\xb_i) = \la \gb(\xb_i; \btheta_0), \btheta^* - \btheta_0\ra, \quad \sqrt{m}\|\btheta^* - \btheta_0\|_2 \leq \sqrt{2\hb^\top\Hb^{-1}\hb} \leq S, \label{lemma:equal_0}
\end{align}
\end{lemma}

\begin{lemma}[Lemma B.4, \citet{zhou2020neural}]\label{lemma:cao_functionvalue}
There exist constants $\{C_i^v\}_{i=1}^3 >0$ such that for any $\delta > 0$, if $\tau$ satisfies that
\begin{align}
     C_1^vm^{-3/2}L^{-3/2}[\log(n_TL^2/\delta)]^{3/2}\leq\tau \leq  C_2^v L^{-6}[\log m]^{-3/2},\notag
\end{align}
then with probability at least $1-\delta$,
% over the random initialization of $\btheta_0$,  
for any $\tilde\btheta$ and $\hat\btheta$ satisfying $ \|\tilde\btheta - \btheta_0\|_2 \leq \tau, \|\hat\btheta - \btheta_0\|_2 \leq \tau$ and $i \in [n_T]$ we have
\begin{align}
    \Big|f(\xb_i; \tilde\btheta) - f(\xb_i;  \hat\btheta) - \la \gb(\xb_i; 
    \hat\btheta),\tilde\btheta - \hat\btheta\ra\Big| \leq C_3^v\tau^{4/3}L^3\sqrt{m \log m}.\notag
\end{align}
\end{lemma}

\begin{lemma}[Lemma B.5, \citet{zhou2020neural}]\label{lemma:cao_gradientdifference}
There exist constants $\{C_i^w\}_{i=1}^3>0$ such that for any $\delta \in (0,1)$, if $\tau$ satisfies that
\begin{align}
     C_1^w m^{-3/2}L^{-3/2}\max \{\log^{-3/2}m, \log^{3/2} (n_T/\delta) \}\leq\tau \leq C_2^w L^{-9/2}\log^{-3}m,\notag
\end{align}
then with probability at least $1-\delta$,
for all $\|\btheta - \btheta_0\|_2 \leq \tau$ and $i \in [n_T]$ we have
\begin{align}
  \| \gb(\xb_i; \btheta) - \gb(\xb_i; \btheta_0)\|_2 \leq C_3^w\sqrt{\log m}\tau^{1/3}L^3\|\gb(\xb_i;  \btheta_0)\|_2.\notag
\end{align}
\end{lemma}

\begin{lemma}[Lemma B.6, \citet{zhou2020neural}]\label{lemma:cao_boundgradient}
There exist constants $\{C_i^z\}_{i=1}^3>0$ such that for any $\delta > 0$, if $\tau$ satisfies that
\begin{align}
 C_1^zm^{-3/2}L^{-3/2}[\log(n_TL^2/\delta)]^{3/2}\leq\tau \leq C_2^z L^{-6}[\log m]^{-3/2},\notag
\end{align}
then with probability at least $1-\delta$,
for any $\|\btheta - \btheta_0\|_2 \leq \tau$ and $i \in[n_T]$ 
we have $\|\gb(\xb_i; \btheta)\|_F\leq C_3^z\sqrt{mL}$.
\end{lemma}

\begin{lemma}[Lemma B.2, \citet{zhou2020neural}]\label{lemma:newboundreference}
There exist constants $\{\bar C_i\}_{i=1}^6>0$ such that for any $\delta > 0$, if for all $t \in [T]$, $\eta$ and $m$ satisfy
\begin{align}
    &\sqrt{2n_t^P/(m\lambda)} \geq \bar C_1m^{-3/2}L^{-3/2}[\log(n_TL^2/\delta)]^{3/2},\notag \\
    &\sqrt{2n_t^P/(m\lambda)} \leq \bar C_2\min\big\{ L^{-6}[\log m]^{-3/2},\big(m(\lambda\eta)^2L^{-6}(n_t^P)^{-1}(\log m)^{-1}\big)^{3/8} \big\},\notag \\
    &\eta \leq \bar C_3(m\lambda + n_t^PmL)^{-1},\notag \\
    &m^{1/6}\geq \bar C_4\sqrt{\log m}L^{7/2}(n_t^P)^{7/6}\lambda^{-7/6}(1+\sqrt{n_t^P/\lambda}),\notag
\end{align}
then with probability at least $1-\delta$,
% over the random initialization of $\btheta_0$, for any $t \in [T]$, 
we have that $\|\btheta_t -\btheta_0\|_2 \leq  \sqrt{2n_t^P/(m\lambda )}$ and
\begin{align}
   \|\btheta_t - \btheta_0 - \hat \bgamma_t\|_2  \leq (1- \eta m \lambda)^{J/2} \sqrt{2n_t^P/(m\lambda)} + m^{-2/3}\sqrt{\log m}L^{7/2}(n_t^P)^{7/6}\lambda^{-7/6}(\bar C_5+ \bar C_6\sqrt{n_t^P/\lambda}).\notag
\end{align}
\end{lemma}

Based on the above definitions and technical lemmas borrowed from previous work, we are now equipped to prove Lemma ~\ref{lemma:neuraluncertain}. We first present the following two lemmas specific to our perturbation based exploration strategy for the regret analysis.

\begin{lemma}
\label{lemma:neuralcb}
For any $t \in [T]$, with $\bar\bomega_t^n$ defined as the solution of the following equation, 
\begin{align}
\label{eq:barloss}
\bar \bomega_t^{n} = \min_{\bomega_t} \sum_{s=1}^{t-1}\sum_{(i, j) \in \Omega_s} &-(y_{i, j}^s + \bar \gamma_{i, j}^{s, (n)})\log \Big ( \sigma(\la\gb_{i, j}^{t, 0}, \bomega_t\ra) \Big ) - (1 - y_{i, j}^s - \bar \gamma_{i, j}^{s, (n)})\log \Big (1 -  \sigma(\la\gb_{i, j}^{t, 0}, \bomega_t\ra) \Big ) \nonumber \\
& + \frac{m\lambda}{2}\sum_{l=1}^p (\eb_l^\top(\btheta-\btheta_0) + \sqrt{\frac{\lambda}{m}}\bar{\gamma}_l^{(n)}\eb_l)^2.
\end{align}
where $\bar \gamma$ represents the expectation of the random variable $\gamma$.
Then, with the pairwise noise $\epsilon_{ij}^s$ satisfying Proposition~\ref{prop:pairwise}, for any document pair $i$ and $j$, with probability at least $1 - \delta$, we have, $|\la\gb_{i, j}^{t,0}, \bar \bomega_t^{n} - \bomega^*\ra| \leq \alpha_t\|\gb_{i, j}^{t, 0}/\sqrt{m}\|_{\Ab_t^{-1}}$,
with $\alpha_t = \frac{1}{c_{\mu}}\big(R\sqrt{\log \frac{\det(\Ab_t^{-1})}{\delta^2\det(\lambda\Ib)}} + \sqrt{\lambda}S\big)$, $\Ab_t = \sum_{s=1}^{t-1}\sum_{(i, j)\in \Omega_s} \gb_{ij}^{s, 0}{\gb_{ij}^{s, 0}}^\top/m + \lambda\Ib$, $S$ is the upper bound of $\sqrt{m} \|\btheta^* - \btheta_0\|_2$ in Lemma~\ref{lemma:equal}.
\end{lemma}

\begin{lemma}
\label{lemma:neuralcb2}
For any $t \in [T]$, with $\hat \bomega_t^{n}$ defined as the solution of the following equation,
\begin{align}
\label{eq:hatloss}
\hat \bomega_t^{n} = \min_{\bomega_t} \sum_{s=1}^{t-1}\sum_{(i, j) \in \Omega_s} &-(y_{i, j}^s + \gamma_{i, j}^{s, (n)})\log \Big ( \sigma(\la\gb_{i, j}^{t, 0}, \bomega_t\ra) \Big ) - (1 - y_{i, j}^s - \gamma_{i, j}^{s, (n)})\log \Big (1 -  \sigma(\la\gb_{i, j}^{t, 0}, \bomega_t\ra) \Big ) \nonumber \\
& + \frac{m\lambda}{2}\sum_{l=1}^p (\eb_l^\top(\btheta-\btheta_0) + \sqrt{\frac{\lambda}{m}}{\gamma}_l^{(n)}\eb_l)^2.
\end{align}
Then, for any pairwise gradient $\gb$, we have $U(\gb) \leq \gb^\top(\hat \bomega_t - \bar \bomega_t) \leq \frac{1}{c_{\mu}}U(\gb)$, and $U(\gb) \sim \cN(0, \nu^2\|\gb/\sqrt{m}\|_{\Ab_t^{-1}}^2)$
\end{lemma}

Based on everything presented above, we are finally ready to provide the detailed proofs of Lemma \ref{lemma:neuraluncertain} as follows.

\begin{proof}

Similar to the proof of Lemma~\ref{lemma:uncertain}, we first analyze for document pair $(i, j)$ satisfying the probability that $i \succ j$ for the given query (e.g., $\sigma(\xijt^\top\btheta^*) > \frac{1}{2}$), the probability that at least one estimation $\sigma(f_{ij}^{t, (n)}) > \frac{1}{2}$ for $n \in [N]$, e.g., $\PP(\max_{n\in[N]} \sigma(f_{ij}^{t, (n)}) > \frac{1}{2}$. Then we have
\begin{align*}
    \PP(\max_{n\in[N]} \sigma (f_{i,j}^{t, (n)}) \geq \frac{1}{2} ) = 1 - \prod_{n=1}^N \PP(\sigma (f_{i,j}^{t, (n)}) < \frac{1}{2})
\end{align*}

For any $n \in [N]$, according to the condition, we have the following bound for $\PP(\sigma (f_{i,j}^{t, (n)}) < \frac{1}{2})$:
\begin{align*}
    \PP(\sigma (f_{i,j}^{t, (n)}) < \frac{1}{2}) &\leq \PP( \sigma(f_{i,j}^{t, (n)}) < \sigma(h_{i,j}) - c_{\mu}\epsilon(m)) \leq \PP(c_{\mu}(f_{i,j}^{t, (n)} - h_{i, j}) < -c_{\mu}\epsilon(m)) \\
    &\leq \PP(\la \gb_{i, j}^0, \hat \bomega_t - \bar \bomega_t\ra  <  \la \gb_{i,j}^0, \bomega^* - \bar \bomega_t  \ra) \leq \PP(\la \gb_{i, j}^0, \hat \bomega_t - \bar \bomega_t\ra  <  \alpha_t\|\gb_{i,j}^0/\sqrt{m}\|_{\Ab_t^{-1}}) \\
    &\leq \PP(U(\gb_{i,j}^0) < \alpha_t\|\gb_{i,j}^0/\sqrt{m}\|_{\Ab_t^{-1}}) \leq 1 - \frac{\exp(-\beta^2)}{4\sqrt{\pi}\beta}
\end{align*}
where $\beta = \frac{\alpha_t}{\nu}$. The first inequality is based on the condition of the lemma, where $\sigma(h_{ij}) - c_{\mu}\epsilon(m) > 1/2$. The second inequality holds with $c_{\mu} = \inf \dot \sigma$. The third inequality holds with the following inequalities.
\begin{align*}
    f(\xb_i; \hat \btheta_{t-1}) - h(\xb_i) = &  f(\xb_i; \hat \btheta_{t-1}) - f(\xb_i; \btheta_0) - \la \gb(\xb_i; \theta_0), \btheta_{t-1} - \btheta_0 \ra + \la \gb(\xb_i; \theta_0), \btheta_{t-1} - \btheta_0 - \hat \bomega_t\ra \\
    &+ \la \gb(\xb_i; \theta_0), \hat \bomega_t - \bar \bomega_t\ra  +  \la \gb(\xb_i; \theta_0), \bar \bomega_t - \bomega^* \ra \\
    \geq& -\epsilon(m)/2 + \la \gb(\xb_i; \theta_0), \hat \bomega_t - \bar \bomega_t\ra  +  \la \gb(\xb_i; \btheta_0), \bar \bomega_t - \bomega^* \ra
\end{align*}
where according to Lemma~\ref{lemma:cao_functionvalue}, Lemma~\ref{lemma:cao_gradientdifference}, Lemma~\ref{lemma:cao_boundgradient} and Lemma~\ref{lemma:newboundreference}, $\epsilon(m)/2 =  C_2^v L^{-6}[\log m]^{-3/2} + C_3^z\sqrt{mL} \cdot ((1- \eta m \lambda)^{J/2} \sqrt{2n_t^P/(m\lambda)} + m^{-2/3}\sqrt{\log m}L^{7/2}(n_t^P)^{7/6}\lambda^{-7/6}(\bar C_5+ \bar C_6\sqrt{n_t^P/\lambda}))$.
The fourth inequality is based on Lemma~\ref{lemma:neuralcb}. The fifth inequality is based on Lemma~\ref{lemma:neuralcb2}. By chaining all the inequalities, we have that with $N \geq \log{\delta}/\log(1 - \exp(-\beta^2)/(4\sqrt{\pi}\beta))$, with probability at least $1 - \delta$, $\max_{n\in[N]} \sigma (f_{i,j}^{t, (n)}) \geq \frac{1}{2}$.

Therefore, under event $E_t$, with the satisfied setting of $N$, based on the definition of $\omega_t^u$ in Section 3.2, we know that for document $i$ and $j$ at round $t$, $(i, j) \in \omega_t^c$ if and only if $\min_{n\in[N]} \sigma\Big(f_{i, j}^{t, (n)}\Big) > \frac{1}{2}$. Then the probability of being uncertain can be bounded as,
\begin{align*}
    \mathbb{P}\big((i, j) \in \omega_t^u\big) = 1 -  \mathbb{P}\big(\min_{n\in[N]}  \hat{\sigma}_t^{(n)} > 1/2\big) = 1 -  \big(\mathbb{P}\big( \hat{\sigma}_t^{(n)} > 1/2\big)\big)^N
\end{align*}
And according to the random matrix theory and Lemma 2 in \citep{jia2021pairrank}, when $t > t^{\prime\prime}$, $\Delta_{\min}/c_{\mu} - CB - \epsilon(m) > 0$ which can be viewed as a constant $c\Delta_{\min}$. Therefore, we have the following inequalities,
\begin{align*}
    \mathbb{P}\big( \sigma(f_{i, j}^{t, (n)}) > 1/2\big) &= \PP(\sigma(f_{i, j}^{t, (n)})  - \sigma(h_{ij}) > 1/2 - \sigma(h_{ij}))) \geq \PP( \sigma(f_{i, j}^{t, (n)}) -\sigma(h_{ij}) > -\Delta_{\min}) \\
    &\geq \PP(f_{i, j}^{t, (n)} - h_{ij} > -\Delta_{\min}/c_{\mu}) \geq \PP(\la\gb_{ij}^{t, 0}, \bar \bomega_t - \hat \bomega_t\ra < \Delta_{\min}/c_{\mu} - \alpha_t\|\gb_{ij}^{t, 0}/\sqrt{m}\|_{\Ab_t^{-1}} - \epsilon(m)) \\
    &\geq \PP(U(\gb_{ij}^{t, 0}) < \Delta_{\min} - \alpha_tc_{\mu}\|\gb_{ij}^{t, 0}/\sqrt{m}\|_{\Ab_t^{-1}} - \epsilon(m)) \geq 1 - \exp(-\frac{c_{\mu}^2c^2\Delta_{\min} ^2}{\nu^2\|\gb_{ij}^{t, 0}/\sqrt{ m}\|_{\bA_t^{-1}}^2})
\end{align*}

Therefore, similar to the analysis for th linear ranker, $\mathbb{P}\big((i, j) \in \omega_t^u\big) \leq \min \Big\{1, \frac{2N\nu^2\|\gb_{ij}^{t, 0}\|_{\bA_t^{-1}}^2}{c_{\mu}^2c^2\Delta_{\min}^2}\Big\}$. This completes the proof.
\end{proof}

\subsubsection{Proof of Lemma~\ref{lemma:neuralcb}}
\label{sec:proofofneuralcb}
\begin{proof}

To obtain the solution $\bar \bomega_t^n$ for Eq~\eqref{eq:barloss}, we first take gradient of the loss function with respect to the parameter. Therefore, $\bar \omega_t^n$ is the solution for the following equation.
\begin{align*}
    \sum_{s=1}^{t-1}\sum_{(i, j)\in\Omega_s} \sigma({\gbijsz}^\top\bomega) - \yijs) \gb_{ij}^{t, 0} + m\lambda\bomega = 0
\end{align*}
Let $q_t(\bomega)= \sum_{s=1}^{t-1}\sum_{(i, j)\in\Omega_s}\sigma({\gbijsz}^\top\bomega)\gbijsz/m + \lambda\bomega$. We know that $q_t(\bar \bomega_t^n) = \sum_{s=1}^{t-1}\sum_{(i, j)\in\Omega_s} \yijs\gbijsz$. Then we can quantify the uncertainty caused by the observation noise for any document pair $i$ and $j$ with the pairwise gradient as $\gb$.
\begin{align*}
    |\gb^\top(\bar\bomega_t - \bomega^*)| &= |\gb^\top\bQ_t^{-1} (q_t(\bar\bomega_t) - q_t(\bomega^*))| \\
    & \leq \frac{1}{c_{\mu}}|\gb^\top\Ab_t^{-1}(q_t(\bar\bomega_t) - q_t(\bomega^*)| \leq \frac{1}{c_{\mu}}\|\gb/\sqrt{m}\|_{\Ab_t^{-1}}\|\sqrt{m}(q_t(\bar\bomega_t) - q_t(\bomega^*))\|_{\Ab_t^{-1}}
\end{align*}
The first equality stands with $\Qb_t = \sum_{s=1}^{t-1}\dot\sigma({\gbijsz}^\top\bomega)\gbijsz{\gbijsz}^\top/m + \lambda\Ib$. According to the definition of $\Ab_t$ and $c_{\mu}$, it is easy to verify that $\Qb_t^{-1} \leq \frac{1}{c_{\mu}}\Ab_t^{-1}$. For the second component of the last equation, we can further bound it by,
\begin{align*}
    \|\sqrt{m}(q_t(\bar\bomega_t) - q_t(\bomega^*))\|_{\Ab_t^{-1}} =& \|\sqrt{m}\big(\sum_{s=1}^{t-1}\xi_{ij}^s\gbijsz/m - \lambda\bomega^*\big)\|_{\Ab_t^{-1}}\\
    \leq& \|\sum_{s=1}^{t-1}\xi_{ij}^s\gbijsz/\sqrt{m} \|_{\Ab_t^{-1}} +  \sqrt{\lambda m}\|\bomega^*\|_2 \leq R\sqrt{\log\frac{\det(\Ab_t)}{\delta^2\det(\lambda\Ib)}} + \sqrt{\lambda}S
\end{align*}
It is easy to have the first and second inequalities. The third inequality holds according to Theorem 2 in \citep{abbasi2011improved}, and the upper bound $S$ defined in Lemma~\ref{lemma:equal}. This completes the proof.
\end{proof}
\subsubsection{Proof of Lemma~\ref{lemma:neuralcb2}}
\begin{proof}
According to the definition of $q_t$ in Section~\ref{sec:proofofneuralcb}, with $\hat \bomega_t^n$ as the solution of Eq~\eqref{eq:hatloss}, we know that $q_t(\hat\bomega_t^n) = \sum_{s=1}^{t-1}\sum_{(i, j) \in \Omega_s} (\yijs + \gamma_{ij}^{s, (n)})\gbijsz/m - \sqrt{\lambda/m}\sum_{l=1}^p \gamma_l^{(n)}\eb_l$. Therefore, for any document pair $i$ and $j$ with the pairwise gradient $\gb$, we know that, 
\begin{align*}
    \gb^\top(\hat \bomega_t - \bar \bomega_t) &= \gb^\top\Qb_t^{-1}(q_t(\hat\bomega_t) - q_t(\bar\bomega_t)) \leq \frac{1}{c_{\mu}} \gb^\top \Ab_t^{-1}(\sum_{s=1}^{t-1}\sum_{(i, j) \in \Omega_s}\gamma_{i,j}^{s, (n)}\gbijsz/m - \sqrt{\lambda/m}\sum_{l=1}^p\gamma_l^{(n)}\eb_l)
\end{align*}
Let $U(\gb) = (\gb/\sqrt{m})^\top \Ab_t^{-1}(\sum_{s=1}^{t-1}\sum_{(i, j) \in \Omega_s}\gamma_{i,j}^{s, (n)}\gbijsz/\sqrt{m} - \sqrt{\lambda}\sum_{l=1}^p\gamma_l^{(n)}\eb_l)$. According to the definition of $\Ab_t$ and $\Qb_t$, it is easy to have $\Qb_t^{-1} \succ \Ab_t^{-1}$. And we can have $\gb^\top(\hat \bomega_t - \bar \bomega_t) \geq U(\gb)$. With $\gamma_{i, j}^{s, (n)}, \gamma_l^{(n)} \sim \cN(0, \nu^2)$, we have $\EE[U(\gb)] = 0$. We have the variance of $U(\gb)$ as,
\begin{align*}
    Var[U(\gb)] = \nu^2\big(\frac{\gb}{\sqrt{m}}^\top\Ab_t^{-1}\big(\sum_{s=1}^{t-1}\sum_{(i, j)\in \Omega_s}\gbijsz{\gbijsz}^\top/m + \lambda\sum_{l=1}^p \eb_l\eb_l^\top\big)\Ab_t^{-1}\frac{\gb}{\sqrt{m}}\big) = \nu^2\|\gb/\sqrt{m}\|_{\Ab_t^{-1}}^2
\end{align*}
This completes the proof.
\end{proof}

\end{document}